\documentclass{article}

\usepackage{amsmath}
\usepackage{PRIMEarxiv}

\usepackage[utf8]{inputenc} 
\usepackage[T1]{fontenc}    
\usepackage{hyperref}       
\usepackage{url}            
\usepackage{booktabs}       
\usepackage{amsfonts}       
\usepackage{nicefrac}       
\usepackage{microtype}      
\usepackage{lipsum}
\usepackage{fancyhdr}       
\usepackage{graphicx}       
\graphicspath{{media/}}     
\RequirePackage{doi}
\usepackage{hyperref}
\usepackage{upgreek}
\title{Dynamic Soaring as a Means to Exceed the Solar~Wind~Speed
\thanks{\textit{\underline{Citation}}: 
\textbf{Larrouturou, M.,  Higgins, A., and Greason, J. \textit{Front. Space Technol.} (2022) DOI:10.3389/frspt.2022.1017442}} 
}
\author{
  Mathias N. Larrouturou, Andrew J. Higgins \\
  Department of Mechanical Engineering \\
  McGill University \\
  Montreal,Quebec, Canada \\
  \texttt{mathias.larrouturou@mail.mcgill.ca} \\
  \texttt{andrew.higgins@mcgill.ca}
   \And
  Jeffrey K. Greason \\
  Tau Zero Foundation\\
  Broomfield, Colorado, USA \\
  Electric Sky Inc., \\
  Midland, Texas, USA \\
  \texttt{jeff@greason.com} \\
}

\begin{document}
\maketitle

\begin{abstract}
A technique by which a spacecraft can interact with flows of ionized gas in space (the solar wind or interstellar medium) in order to be accelerated to velocities greater than the flow velocity is explored. Inspired by the dynamic soaring maneuvers performed by sea birds and gliders in which differences in wind speed are exploited to gain velocity, in the proposed technique a lift-generating spacecraft circles between regions of the heliosphere that have different wind speeds, gaining energy in the process without the use of propellant and only modest onboard power requirements. In the simplest analysis, the spacecraft motion can be modeled as a series of elastic collisions between regions of the medium moving at different speeds. More detailed models of the spacecraft trajectory are developed to predict the potential velocity gains and the maximum velocity that may be achieved in terms of the lift-to-drag ratio of the vehicle. A lift-generating mechanism is proposed in which power is extracted from the flow over the vehicle in the flight direction and then used to accelerate the surrounding medium in the transverse direction, generating lift (i.e., a force perpendicular to the flow). Large values of lift-to-drag ratio are shown to be possible in the case where a small transverse velocity is imparted over a large area of interaction. The requirement for a large interaction area in the extremely low density of the heliosphere precludes the use of a physical wing, but the use of plasma waves generated by a compact, directional antenna to impart momentum on the surrounding medium is feasible, with the excitation of R-waves, X-waves, Alfven waves, and magnetosonic waves appearing as promising candidates. A conceptual mission is defined in which dynamic soaring is performed on the termination shock of the heliosphere, enabling a spacecraft to reach speeds approaching 2\% of $c$ within two and a half years of launch without the expenditure of propellant. The technique may comprise the first stage for a multistage mission to achieve true interstellar flight to other solar systems.
\end{abstract}


\section{Introduction}
\label{sec:intro}
The extreme energy requirements necessary to achieve interstellar flight comprise a formidable challenge. By examining the power available to civilization and the fraction that might be devoted to spaceflight, it can be argued that launch of a Voyager-class spacecraft to $\upalpha$-Centauri with transit times comparable to a human lifetime will not be feasible until the 25th~century \cite{Millis2010}. This consideration suggests that exploring sources of energy available in space that can be exploited for propulsion is worthwhile.

Solar sails are the first example of a propulsion technology that utilizes the freely available photons emanating from the sun, but even the most extreme solar sailing---launched from near the sun using the highest temperature materials with the lowest areal density (e.g., aerographite)---would only be capable of achieving 2\% of $c$ \cite{Heller2020}; more conventional solar sails are limited to less than 0.5\% of $c$ \cite{Davoyan2021}. Recently, Lingam and Loeb \cite{Lingam2020} have examined astrophysical objects (e.g., massive stars, supernovae, etc.) that would permit a radiation-pushed light sail to achieve velocities of 10\% of $c$ or greater, but this still leaves the problem of how human technology originating from the solar system can achieve interstellar flight.

Over the last three decades, the idea of utilizing the solar wind of charged particles streaming from the sun for propulsion has attracted attention. Although the available power-per-unit-area of the solar wind ($\frac{1}{2} \rho_\infty v_\mathrm{\infty}^3$) in the vicinity of the earth is six orders of magnitude less than the solar photon flux, the fact that the solar wind is comprised of charged particles means that it may be interacted with electromagnetically rather than with a solid surface, enabling a relatively small, low-mass structure to interact with an enormous area of wind.

Concepts for generating force on a spacecraft via interaction with the solar wind include the magnetic sail, the electric sail, and the plasma magnet. The magnetic sail (MagSail) as originally proposed by Zubrin and Andrews \cite{Zubrin1991, Zubrin2000} would consist of a loop of superconducting cable that generates an artificial magnetosphere, deflecting the stream of charged particles in the solar wind and imparting a respective reaction force on the cable.  The electric sail (E-sail), as proposed by Janhunen \cite{Janhunen2004, Giovanni2008}, dispenses with the need for a superconducting cable and instead relies on wires charged to  a high voltage to deflect the charged particles of the solar wind. The plasma magnet \cite{Slough2005, Slough2007} would use a polyphase antenna onboard the spacecraft to drive currents in the surrounding medium, creating a magnetic structure that would inflate via self-repulsion until the magnetic pressure is balanced by the dynamic pressure of the impinging solar wind. The plasma magnet appears particularly encouraging in that it is able to interact with an enormous volume of the solar wind (e.g., tens to hundreds of km in extent) while only requiring a small antenna (e.g., meters) with modest power requirements. See \cite{Djojodihardjo2018} for a recent review of magnetic sail concepts. 
Magnetic sails are capable of only producing modest amounts of lift ($\frac{L}{D} \approx 0.3$ \cite{Zubrin1991, Perakis2020}); they are predominately drag devices, more similar to parachutes than sails, which could be dragged up to the speed of the solar wind ($\approx700$~km/s).

A technique does exist that permits a vehicle interacting with the wind to exceed the velocity of the wind:  \emph{dynamic soaring}. In dynamic soaring as practiced terrestrially, a lift-generating vehicle executes a maneuver that exploits the difference in wind speeds between two different regions of air, for example, the wind blowing over a hilltop and the quiescent air on the leeward side of the hill. This technique, as invoked by seabirds, was noted as early as c. 1513--1515 in the notebooks of Leonardo da Vinci \cite{ Richardson2019}. The first quantitative, physics-based explanation of this behavior by birds was advanced by Lord Rayleigh in 1883 \cite{Rayleigh1883}:  In essence, the vehicle (or bird) executes an elastic collision when entering the moving stream of air via a low-drag banking maneuver. As the vehicle re-enters the quiescent air, it has gained \emph{twice} the velocity of the wind stream. By then banking in the quiescent air, the vehicle can re-enter the wind stream and increase its velocity again, repeating the maneuver over and over until drag losses counteract the velocity gains and a maximum velocity is achieved. Recently, remote control glider enthusiasts have achieved remarkable velocities exceeding 850~km/hr---approximately 10 times the speed of the wind---by invoking this technique with gliders that have no onboard propulsion \cite{Lisenby2017, Blain2021}.

The possibility of exploiting dynamic soaring for spaceflight using a magnetohydrodynamic wing with a large value of lift-to-drag ratio was first suggested by Birch \cite{Birch1989}, who proposed a vehicle interacting with the turbulent clouds of the interstellar medium (ISM) in order to gain velocity. Interestingly, charged particles bouncing between turbulent clouds in the ISM was suggested by Fermi as a potential source of galactic cosmic rays (GCRs) in 1949 \cite{Fermi1949}, a mechanism now called second-order Fermi acceleration. It has since been shown that acceleration of charged particles via astrophysical shock waves (such as originating from supernovae) provides a more likely process for producing nonthermal particles and explains the observed power-law relation of GCRs. The shock-wave mechanism, suggested by Fermi in a follow-up publication \cite{Fermi1954} and now termed first-order Fermi acceleration, is often likened to an elastic ball bouncing between two approaching trains (with the trains representing the flows upstream and downstream of the shock wave):  The ball (or charged particle) gains momentum with each bounce across the shock proportional to the difference in flow velocities \cite{Spurio2015}. This mechanism of charged particle acceleration can be considered a type of dynamic soaring which occurs in an astrophysical context.

In this paper, a new concept for spacecraft propulsion that invokes dynamic soaring is proposed. In this concept, lift is generated by extracting power in one direction (in the direction of the medium blowing over the spacecraft) and accelerating flow in the other (perpendicular) direction. This approach builds upon the concept of a propulsive drive that is powered by external dynamic pressure (the so-called \emph{q-drive} \cite{Greason2019}), however, in the present concept, no onboard reaction mass is used. By using the external power generation to accelerate matter available in the solar wind perpendicular to the flow over the vehicle, lift is generated that is greater in magnitude than the drag generated by the power-extraction process.  The result is a type of lift-generating wing, but without a physical structure. In Section~2, the operating principles of this lift-generating mechanism are developed in detail. In Section~3, potential mission concepts are developed utilizing regions of high wind shear available in the solar system, namely, the interface between the fast (polar) and slow (equatorial) solar wind and the termination shock where the solar wind reverts from supersonic to subsonic flow, to reach velocities of $\approx\!2$\% of $c$. In Section~4, mechanisms for extracting power from the flow and ways of using power to impart transverse motion onto the flow are explored, with special attention devoted to plasma waves as a particularly promising means of lift generation. Finally, in Section~5, a candidate mission is proposed and the potential for synergistic combinations of the dynamic soaring technique with other externally powered propulsive devices to achieve even greater velocities is discussed.

\begin{figure}
    \centering
    \includegraphics[scale=0.45]{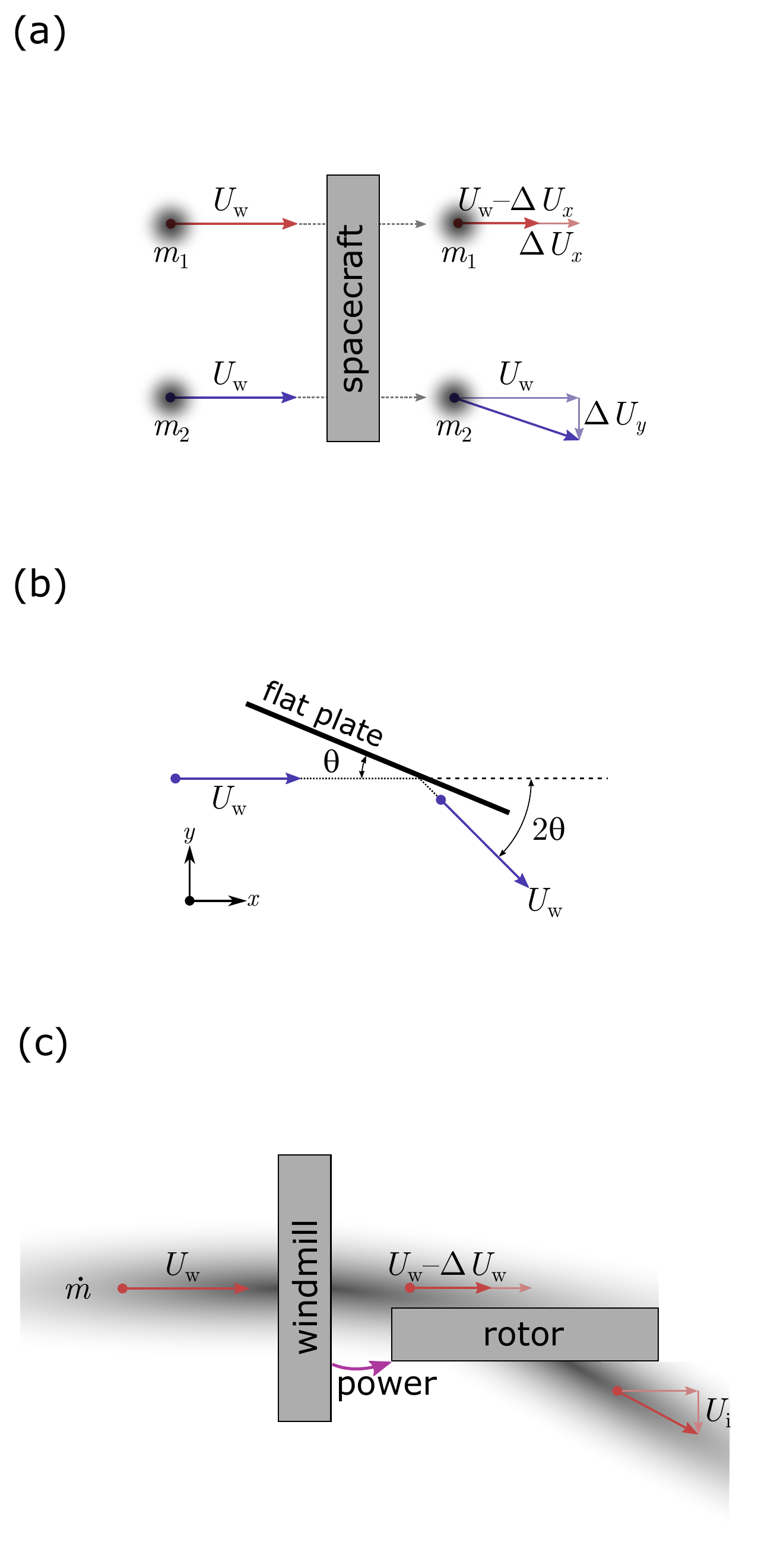}
    \caption{Schematic depiction of spacecraft interacting with two mass streams}
    \label{fig:Intro01}
\end{figure}

\section{Generating a high lift-to-drag ratio}
\label{sec:highLoverD}

In considering how to achieve lifting flight in the diffuse interplanetary or interstellar medium, it is necessary to examine more closely what we mean by \emph{lift}.  Canonically, in aerodynamics, drag is the aerodynamic force parallel to the apparent wind (that is, the velocity of the medium in the frame of reference of the vehicle), while lift is an aerodynamic force perpendicular to the apparent wind \cite{Anderson1991}.

The mechanism by which lift can be generated by interacting with a stream of particles blowing past a vehicle is illustrated in Fig.~\ref{fig:Intro01}(a), which is formulated in a vehicle-fixed reference frame. The vehicle and its interaction with streams of particles is represented by a control volume. Power is extracted from stream~1, decreasing the velocity by $\Delta U_x$. This power is then used to accelerate stream~2 by increasing its velocity component by $\Delta U_y$ in the direction perpendicular to the original stream.  If energy is conserved in this interaction,

\begin{equation} \label{eqn:eqat1}
    \frac{1}{2}\dot{m}_2\left(U_{\mathrm{w}}^2+\left(\Delta U_{y}^2\right)\right)-\frac{1}{2} \dot{m}_2\,U_{\mathrm{w}}^2=\frac{1}{2} \dot{m}_1\,U_{\mathrm{w}}^2 - \frac{1}{2} \dot{m}_1\,\left(U_{\mathrm{w}}-\Delta U_{x}\right)^2
\end{equation}

\begin{equation} \label{eqn:eqat2} 
    \Delta U_{y}  = 
    \sqrt{\left(2\,U_\mathrm{w}\,\Delta U_{x} - \left(\Delta U_{w}\right)^{2}\right)\frac{\dot{m_\mathrm{1}}}{\dot{m_\mathrm{2}}}}
    \approx 
    \sqrt{2\,U_\mathrm{w}\,\Delta U_{x}
    \frac{\dot{m_\mathrm{1}}}{\dot{m_\mathrm{2}}}}.
\end{equation}
\newline
\noindent
In the limit that the mass flow becomes large and $\dot{m}_1=\dot{m}_2$, the lift-to-drag ratio can now be derived. 
\begin{equation} \label{eqn:eqat3} 
    \frac{L}{D}=\frac{\dot{m}_2\,\Delta U_{y}}{\dot{m}_1\,\Delta U_{x}}
    =2\frac{U_\mathrm{w}}{\Delta U_{y}}.
\end{equation}
\noindent
This result demonstrates the general principle of lift generation:  If the change in velocity in the perpendicular direction ($\Delta U_y$) is made small in comparison to the wind speed, then a large lift-to-drag ratio can be obtained.
These results can be obtained by having the particles interact with an inclined surface via collisions, as depicted in Fig.~\ref{fig:Intro01}(b) \cite{Anderson1991}.  If the particles undergo an energy conserving collision (i.e., an elastic collision), then the collision is specular and the momentum imparted on the surface can be resolved into lift and drag components:

\begin{equation} \label{eqn:eqat4} 
    \frac{L}{D}=\frac{U_\mathrm{w}\sin{\theta}}{U_\mathrm{w}-U_\mathrm{w}\cos{\theta}}
    =    \frac{\sin{\theta}}{1-\cos{\theta}} 
    \approx \frac{2}{\theta}
\end{equation}

\begin{equation} \label{eqn:eqat5}
    \Leftrightarrow \frac{L}{D}=
    \frac{\frac{\Delta U_{y}}{U_\mathrm{w}}}{1-\frac{U_\mathrm{w}-\Delta U_{x}}{U_\mathrm{w}}}
    =2\frac{U_\mathrm{w}}{\Delta U_{y}}
\end{equation}
\noindent
which is the same result as Eq.~\ref{eqn:eqat3} above. This result is also similar to the classic results of a flat plate at an angle of attack in Newtonian hypersonic flow ($ L/D \sim \frac{1}{\tan(\alpha)}\sim \frac{1}{\alpha}$), but differs by a factor of two due to the specular reflection (rather than particles sliding along the plate, as in Newtonian flow). This result clarifies that power extraction in the flow direction and using that power to accelerate flow in the transverse direction is a type of aerodynamic lift, similar to a flat plate in hypersonic flow. For the purposes of the concept discussed in this paper, however, it is useful to continue to treat the power extraction and transverse acceleration as separate processes. Although lift and drag are usually discussed in a continuum regime, the continuum assumption was not invoked in this particle-based model. Similar to how drag devices such as the MagSail \cite{Zubrin1991, Zubrin2000} and the E-sail \cite{Janhunen2004, Giovanni2008} are not required to operate in the continuum regime in order to produce drag, the interaction mechanism described earlier does not require continuum flow to produce lift.

In conventional hypersonic lift analysis of flat plates, there is a limit to the lift-to-drag ratio that is achievable due to skin friction of the hypersonic flow over the physical plate. In the case that we are discussing with interactions in space plasmas, there is no physical plate (i.e., wing) and plasmas in space are usually collisionless. Therefore, the limits on achievable lift-to-drag ratio in hypersonic flat plate theory do not apply to the present case.

A third way of of conceptualizing a lift-generating glider is to envision a wind energy extraction device, such as a rotary wind turbine or windmill aligned with the flow over the vehicle, and a propeller or rotor aligned with the lift axis, as shown in Fig.~\ref{fig:Intro01}(c). If the profile of the rotor blades and the rotor hub is negligibly small, so that they are not creating significant shocks in the wind, and if skin friction is negligible, then:	
 \begin{equation} \label{eqn:eqat6}
     \frac{1}{2}\dot{m}\ U_{\mathrm{w}}^2 -\frac{1}{2}\dot{m}  \Big( U_{\mathrm{w}}- \Delta U_{\mathrm{w}} \Big) ^2
     = \frac{1}{2} \dot{m}\,U_{\mathrm{i}}^2 
 \end{equation}
 
 \begin{equation} \label{eqn:eqat7}
     \Rightarrow
     U_{\mathrm{i}}
     = 
     \sqrt{2\,v_{\mathrm{w}}\,\Delta U_{\mathrm{w}}-\Delta U_{\mathrm{w}}^2}
     \approx
      \sqrt{2\,v_{\mathrm{w}}\,\Delta U_{\mathrm{w}}}
 \end{equation}
 
 \begin{equation} \label{eqn:eqat8}
     \Rightarrow
     \frac{L}{D} 
     =
     \frac{\dot{m}\,v_{\mathrm{i}}}{\dot{m}\,\Delta U_{\mathrm{w}}}
     =
     \frac{U_{\mathrm{i}}}{\Delta U_{\mathrm{w}}}
     =
     2\frac{v_\mathrm{w}}{U_{\mathrm{i}}}.
 \end{equation}
 \noindent
 These results are familiar in aerodynamics as the \emph{propeller equation} \cite{Anderson1991} in the limit where the mass flows are large, giving:	
 
\begin{equation} \label{eqn:eqat9}
    P_{\mathrm{windmill}}= D\,U_{\mathrm{w}}
\end{equation}
\begin{equation} \label{eqn:eqat10}
    P_{\mathrm{rotor}}= \frac{1}{2} L\,U_{\mathrm{i}}.
\end{equation}
\noindent
Equating ~\ref{eqn:eqat9} and \ref{eqn:eqat10}, and rearranging gives:

 \begin{equation} \label{eqn:eqat11}
     \frac{L}{D} 
     \approx
     2\frac{U_{\mathrm{w}}}{U_{\mathrm{i}}}=2\frac{U_{\mathrm{w}}}{U_{y}}.
 \end{equation}
\noindent
Therefore, the analysis based on the propeller (or rotor) equation produces a result identical to the particle-based results of Eqs.~\ref{eqn:eqat3} and \ref{eqn:eqat5}. In general, the extraction of power from the wind (the \emph{windmill}) and acceleration of the transverse flow cannot be accomplished without losses.  An efficiency of the power transfer can be introduced:
\begin{equation} \label{eqn:eqatpower}
     P_{\mathrm{lift}} = \eta \, P_{\mathrm{windmill}}
\end{equation}
\begin{equation} \label{eqn:eqat13}
    \Rightarrow
    \frac{L}{D}=2 \, \eta\,\frac{U_\mathrm{w}}{\Delta U_{y}}
\end{equation}
\begin{figure}[h]
    \centering
    \includegraphics[scale=0.65]{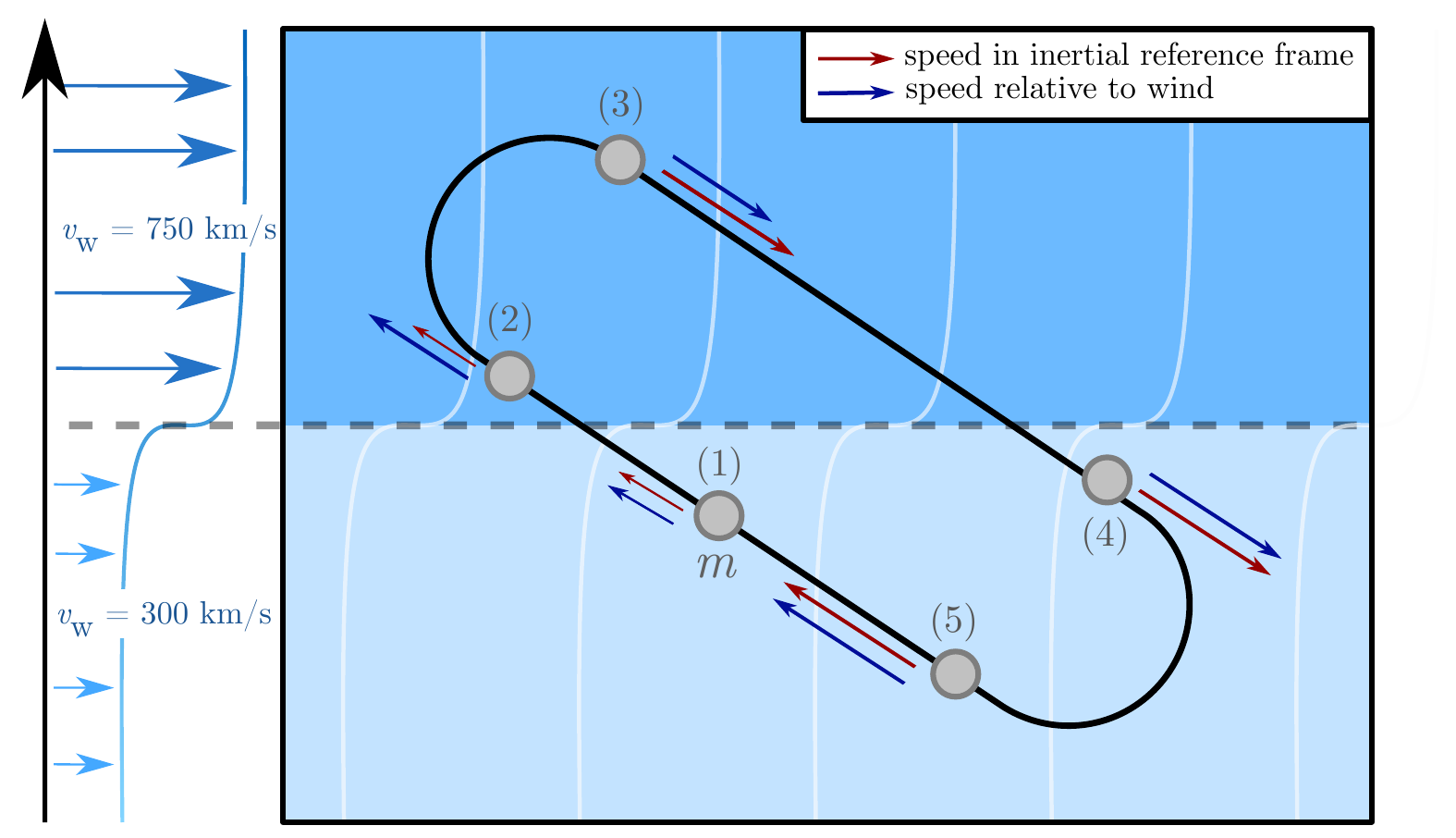}
    \caption{The trajectory of a vehicle performing dynamic soaring in wind shear on the slow and fast solar wind}
    \label{fig:DSTS}
\end{figure}

\section{Dynamic soaring}

Albatrosses can stay airborne for months without flapping their wings \cite{gao2015, richardson2011}. To achieve this feat, albatrosses use a technique called dynamic soaring. They oscillate between a fast wind region and a slower wind region, extracting energy from the wind. Through this process they are able to gain both altitude and speed. Through the gain in altitude, birds are able to store energy and reduce their velocity in the faster wind region, increasing the maximum achievable speed. However, using potential energy this way is not necessary for the dynamic soaring maneuver to be viable. In space soaring, the vehicle soars faster than the orbital velocity required to stay in orbit and the accelerations produced by the propulsion system are two orders of magnitude greater than the gravitational acceleration of the Sun. Therefore, gravity does not play a role in space soaring.

The simplest two-dimensional dynamic soaring trajectory across a wind velocity gradient is illustrated in Fig.~\ref{fig:DSTS}. As shown, a vehicle enters the stronger wind region experiencing a strengthening head wind. A banking maneuver is then performed, increasing the ground speed. The vehicle enters the weaker wind region with a weakening tail wind, once again gaining airspeed. A second banking maneuver is executed in the slower wind region. By entering with a strengthening head wind and leaving with a weakening tail wind, the spacecraft gains twice the difference in wind speed between the two regions. Due to the low drag associated with each banking maneuver, the entire trajectory can be seen as a sequence of elastic collisions between the vehicle and the two regions of different wind speed. The maneuver can be repeated until the losses from drag equal the velocity gain.

\subsection{Analytical model for dynamic soaring}

\begin{figure}
    \centering
    \includegraphics[scale=0.65]{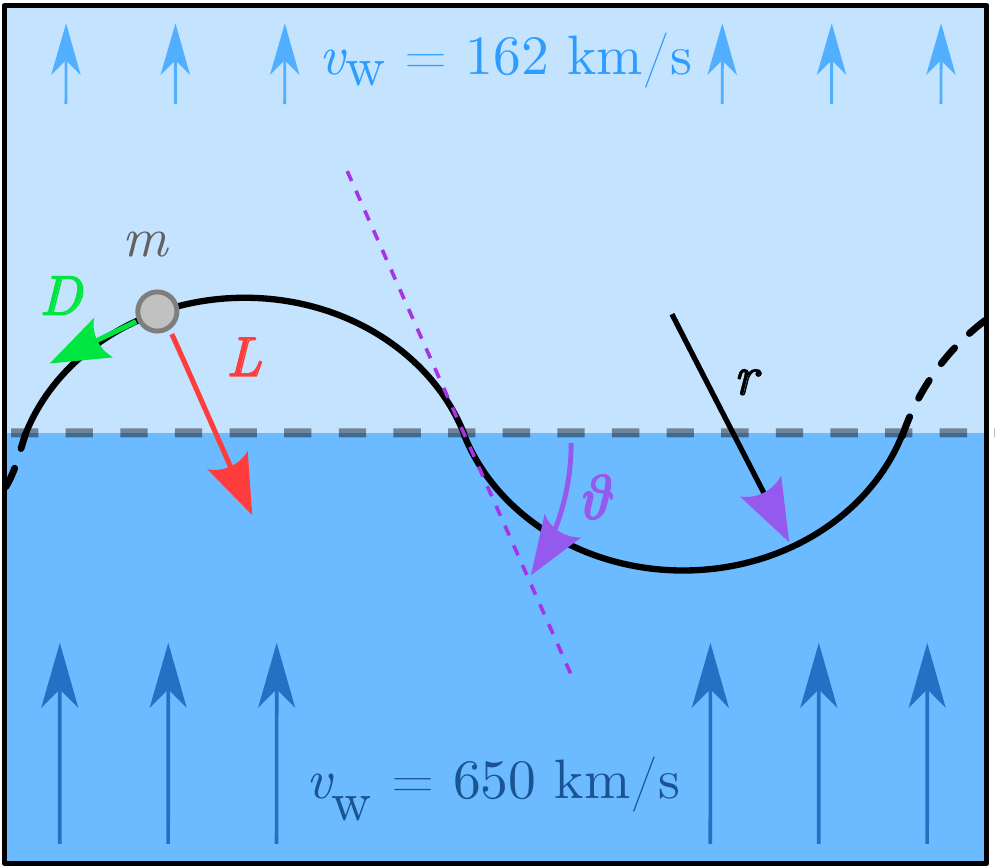}
    \caption{Two-Dimensional Dynamic Soaring trajectory on a normal shock}
    \label{fig:DSTS2}
\end{figure}
Dynamic soaring is a power extraction process; it can therefore be understood by looking at the transfer of energy between the medium and the spacecraft. In our case, the mechanical energy of the system is simply the spacecraft kinetic energy. The speed and acceleration of the spacecraft are modeled in the reference frame of the wind, while the position of the spacecraft is modeled in the fixed reference frame. 
The kinetic energy of the spacecraft relative to the wind can be written as,

\begin{equation} \label{eqn:eqat13}
    E_\mathrm{kin}=
    \frac{m}{2}\left( \boldsymbol{U} \cdot\boldsymbol{U} \right) 
\end{equation}

\begin{equation} \label{eqn:eqat14}
    \Rightarrow 
    \frac{\mathrm{d}E_\mathrm{kin}}{\mathrm{d}t}=m\left(\boldsymbol{U}\cdot\frac{\mathrm{d}\boldsymbol{U}}{\mathrm{d}t}\right).
\end{equation}

\noindent
By definition $\boldsymbol{U}=\boldsymbol{v}-\boldsymbol{v_\mathrm{w}}$  
 $\Rightarrow \frac{\mathrm{d}\boldsymbol{U}}{\mathrm{d}t}=\frac{\mathrm{d}\boldsymbol{v}}{\mathrm{d}t}-\frac{\mathrm{d}\boldsymbol{v_\mathrm{w}}}{\mathrm{d}t}$
 $\Rightarrow \frac{\mathrm{d}\boldsymbol{U}}{\mathrm{d}t}=\frac{\mathrm{d}^2\boldsymbol{x}}{\mathrm{d}t^2}-\frac{\mathrm{d}\boldsymbol{v_\mathrm{w}}}{\mathrm{d}t}$

\begin{equation} \label{eqn:eqat15}
    \frac{\mathrm{d}E_\mathrm{kin}}{\mathrm{d}t}=m\left(\boldsymbol{U}\cdot \left(\frac{\mathrm{d}^2\boldsymbol{x}}{\mathrm{d}t^2}-\frac{\mathrm{d}\boldsymbol{v_\mathrm{w}}}{\mathrm{d}t} \right) \right)
\end{equation}

\begin{equation} \label{eqn:eqat16}
    \Rightarrow
    \frac{\mathrm{d}E_\mathrm{kin}}{\mathrm{d}t}=\boldsymbol{U}\cdot\left(m\, \frac{\mathrm{d}^2\boldsymbol{x}}{\mathrm{d}t^2}\right)-\boldsymbol{U}\cdot\left(m\,\frac{\mathrm{d}\boldsymbol{v_\mathrm{w}}}{\mathrm{d}t} \right).
\end{equation}
\noindent
The first term in brackets $(m \frac{\mathrm{d}^2 x}{\mathrm{d}t^2})$  represents the total force acting on the system. In our case, the only forces acting on the system are the lift and the drag. Lift is perpendicular to the apparent wind, hence, $\boldsymbol{U}\cdot\boldsymbol{L}=0$. Equation~\ref{eqn:eqat16} can be further expanded by developing the dot products into their scalar form 

\begin{equation} \label{eqn:eqat17}
    \Rightarrow
    \frac{\mathrm{d}E_\mathrm{kin}}{\mathrm{d}t}=-U{\cdot D}-U\left(m\,\frac{\mathrm{d}v_\mathrm{w}}{\mathrm{d}t}\,\cos{\phi}\right)
\end{equation}
with $\phi$ being the angle between the wind gradient and the airspeed vector.
This equation describes all energy transfer between the medium and the system. The first term represents the losses due to drag and the second term is usually referred to as the dynamic soaring force,\, $F_\mathrm{dyn}=m\,\frac{\mathrm{d}\boldsymbol{v_\mathrm{w}}}{\mathrm{d}t}\,\cos{\phi}$.

For the system to extract energy, the dot product in the last term of Eq.~\ref{eqn:eqat16} must be negative. Therefore, the spacecraft must soar in a strengthening head wind or in a weakening tail wind. If the spacecraft follows a trajectory as shown in Fig.~\ref{fig:DSTS}, the velocity gain over one cycle from dynamic soaring can be defined as

\begin{equation} \label{eqn:eqat18}
    \Delta v_\mathrm{dyn} = \Delta U= 2\,\Delta v_\mathrm{w}\,\cos{\phi}.
\end{equation}
The optimum value for the $\phi$ is derived in the following sections.

\subsubsection{Simple analytical setup}

Consider two wind layers, one moving at speed $v_{\mathrm{\infty}}$ and the other one at rest, with an instantaneous jump in wind speed at the interface. The motion of the spacecraft will be limited to the vertical plane and will be modeled entirely in a fixed reference frame.  The wind is moving in the $x$-direction while the wind gradient is in the $y$-direction. The spacecraft is assumed to follow an arc of constant curvature as shown in Fig.~\ref{fig:DSTS2}, and crosses the interface with a constant angle, $\theta$.
Equation \ref{eqn:eqat18} can be rewritten as $\Delta u = \Delta v_\mathrm{w}\sin{\theta}$.

The lift can be related to the centripetal acceleration required to follow the trajectory,
\begin{equation} \label{eqn:eqat19}
    \boldsymbol{a}_{\mathrm{{centr}}}=\boldsymbol{L} \Rightarrow 
    m \frac{v_{\mathrm{s}}^2}{r}=\frac{1}{2}\,\rho_{\mathrm{\infty}}\, C_{\mathrm{L}}\, S_\mathrm{ref} \left(v_{\mathrm{\infty}} -v_{\mathrm{s}} \right)^2 
\end{equation}
\begin{equation} \label{eqn:eqat20}
   \Rightarrow r=\frac{2m}{\rho_{\mathrm{\infty}}\, C_{\mathrm{L}}\, S_\mathrm{ref} \left(\frac{v_{\mathrm{\infty}}}{v_{\mathrm{s}}}-1 \right)^2}\approx \frac{2\,\beta}{\rho_{\mathrm{{\infty}}}\frac{L}{D}}
\end{equation}
with $\beta=\frac{m}{C_{\mathrm{D}}S}$ as the ballistic coefficient of the spacecraft. In the limit of $v_{\mathrm{\infty}} << v_{\mathrm{s}}$, the final expression of the radius of the trajectory followed by the spacecraft is solely determined by the spacecraft properties (i.e., the ballistic coefficient and the lift-to-drag ratio) and the density of the medium.

Over each cycle the spacecraft gains twice the speed of the wind, from which the losses due to drag have to be subtracted. We will consider drag as being constant over each circular trajectory, therefore, the velocity losses can be expressed as

\begin{equation} \label{eqn:eqat21}
    \Delta v_{\mathrm{{drag}}}= D \frac{\Delta t}{m}
    =\frac{1}{2}\,\rho_{\mathrm{{\infty}}}\,C_{\mathrm{D}}\, S_\mathrm{ref} \left(v_{\mathrm{{\infty}}}-v_{\mathrm{s}}\right)^2 \frac{4\,\theta \,r}{v_{\mathrm{s}}\,m}
\end{equation}
Taking the limit as $v_{\inf} << v_\mathrm{s}$, 
\begin{equation} \label{eqn:eqatv_drag}
    v_{\mathrm{{drag}}}= \frac{2\,\theta\,v_{\mathrm{s}}}{\frac{L}{D}}.
\end{equation}
\noindent
Equations \ref{eqn:eqat18} and \ref{eqn:eqatv_drag} have competing effects, dictating the optimal trajectories that must be followed by the spacecraft.

\subsubsection{Maximum speed gain per cycle}

By considering the velocity gained through the dynamic soaring maneuver and the losses due to drag, an analytic solution of the maximum speed gain per cycle can be derived.

\begin{equation} \label{eqn:eqat22}
    \Delta v_{\mathrm{cycle}}=\Delta v_{\mathrm{dyn}}-\Delta v_{\mathrm{drag}} 
    = 2 \, v_{\mathrm{\infty}}\,\sin{\theta}-\frac{2\,\theta\, v_{\mathrm{s}}}{\frac{L}{D}}-
    \frac{2\,\theta\left( v_{\mathrm{s}}+2\, v_{\mathrm{\infty}}\right)}{\frac{L}{D}}
\end{equation}
\begin{equation} \label{eqn:eqat23}
    \Rightarrow \Delta v_{\mathrm{cycle}} = 2 \, v_{\mathrm{\infty}}\,\sin{\theta}-\frac{4\,\theta\left(v_{\mathrm{s}}+v_{\mathrm{\infty}}\right)}{\frac{L}{D}}.
\end{equation}

\subsubsection{Maximum speed limit}
The maximum speed is reached by the spacecraft when the velocity gained through the dynamic soaring force equals the velocity loss due to drag, or in other words, when the velocity gained per cycle approaches zero.
From the two-wind-layer model, an analytic solution can be derived for the maximum reachable speed. 

\begin{equation} \label{eqn:eqat24}
    \Delta v_{\mathrm{dyn}}=\Delta v_{\mathrm{drag}} 
    \Rightarrow 
2 v_{\mathrm{\infty}}\sin{\theta}
    =\frac{2\,\theta\, v_{\mathrm{s}}}{\frac{L}{D}}+\frac{2\,\theta\left( v_{\mathrm{s}}+2\,v_{\mathrm{\infty}}\right)}{\frac{L}{D}}
\end{equation}
Isolating for the speed of the spacecraft yields, 
\begin{equation} \label{eqn:eqat25}
    v_{\mathrm{s, max}}=\left(v_{\mathrm{\infty}}\sin{\theta}-\frac{2\,\theta\sin{\theta}}{\frac{L}{D}}v_{\mathrm{\infty}}\right)\left(\frac{\frac{L}{D}}{2\,\theta}\right)
    \approx \frac{v_{\mathrm{\infty}}}{2} \left(\frac{L}{D} \frac{\sin{\theta}}{\theta}\right).
\end{equation}
In the limit as $\theta$ tends to $0$, 
\begin{equation} \label{eqn:eqat26}
    v_{\mathrm{s, max}}
    =
    \frac{v_{\mathrm{\infty}}}{2} \left(\frac{L}{D}\right).
\end{equation}
By considering a more general case where both wind layers have a non-zero wind speed, a similar result can be derived. The spacecraft still follows arcs of constant curvature, crossing the interface at a fixed angle.  Let $\Delta v_\mathrm{w}=v_{\mathrm{\infty, 1}}-v_{\mathrm{\infty, 2}}$,

\begin{equation} \label{eqn:eqat27}
    \Delta v_{\mathrm{cycle}}=\Delta v_{\mathrm{dyn}}-\Delta v_{\mathrm{drag}} 
    = 2 \Delta v_\mathrm{w}\,\sin{\theta}-\frac{2\,\theta\, v_{\mathrm{s}}}{\frac{L}{D}}-
    \frac{2\,\theta\left( v_{\mathrm{s}}+2\, \Delta v_\mathrm{w}\right)}{\frac{L}{D}}
\end{equation}
\begin{equation} \label{eqn:eqat28}
    \Rightarrow \Delta v_{\mathrm{cycle}} = 2 \Delta v_\mathrm{w}\,\sin{\theta}-\frac{4\,\theta\left(v_{\mathrm{s}}+\Delta v_\mathrm{w}\right)}{\frac{L}{D}}.
\end{equation}
\noindent
Equating $\Delta v_{\mathrm{cycle}}$ to zero yields, 

\begin{equation} \label{eqn:eqat29}
    v_{\mathrm{s, max}}=\left(\Delta v_\mathrm{w}\sin{\theta}-\frac{2\,\theta\sin{\theta}}{\frac{L}{D}}\Delta v_\mathrm{w}\right)\left(\frac{\frac{L}{D}}{2\theta}\right)
\end{equation}
\begin{equation} \label{eqn:eqat30}
     v_{\mathrm{s, max}}
     \approx \frac{\Delta v_\mathrm{w}}{2} \left(\frac{L}{D} \frac{\sin{\theta}}{\theta}\right).
\end{equation}
As $\theta$ tends to $0$, 
\begin{equation} \label{eqn:eqat31}
    v_{\mathrm{s, max}}\approx \frac{\Delta v_\mathrm{w}}{2} \left(\frac{L}{D}\right).
\end{equation}

\noindent
From Eq. \ref{eqn:eqat31} we can see that as the angle decreases, the maximum reachable velocity increases. The greatest maximum speed arises as the angle tends to 0$^\circ$ and the spacecraft glances off the interface.

\subsection{Possible trajectories}

Several structures in the solar system offer wind gradients large enough for dynamic soaring maneuvers to extract energy. Such structures include but are not limited to: the termination shock \cite{balogh2013, Mccomas2019}, the heliopause  \cite{balogh2013, Mccomas2019, krimigis2011}, the slow and fast solar wind \cite{Miralles2011, verscharen2019}, and the boundary of the planetary magnetosphere\cite{Zwan1976, Spreiter1980, Slavin1981}.  While the density of these structures varies, analysis of drag devices such as the plasma magnet have shown that the extend of the artificially generated magnetosphere around the vehicle naturally expands as the surrounding density decreases\cite{Greason2019, freeze2021wind, Slough2005, Slough2007}.Specifically, the magnetic structure around the spacecraft will expand until the magnetic pressure matches the dynamic pressure of the solar wind. This effect makes devices such as the plasma magnet nearly constant drag as they move outward from the sun. For the purposes of the analysis in this paper, we have adopted constant values of drag and, since the lift generated derives from the oscillation of the drag device through the plasma, constant values of lift as well.

\subsubsection{Termination shock and Heliopause}
The termination shock is the region where the solar wind goes from supersonic to subsonic. The solar wind speed decreases from 650~km/s to 200~km/s. As will be explained later on, the spacecraft interacts with the charged particles within the medium. Therefore, we are considering the velocity of the flow of charged particles (i.e., protons and electrons) and not considering the velocity of the neutral particles. The solar wind speed at high latitude, just upstream of the termination shock, is expected to be $~650~\mathrm{km/s}$ \cite{usmanov2006}. The velocity downstream of the shock is uncertain; we expect on theoretical grounds that it is less than 200~km/s (our simulations have assumed 162~km/s). The velocity after the termination shock continues to slow until it reaches zero outflow speed near the heliopause  \cite{balogh2013, Mccomas2019, krimigis2011}. This second structure offers a slower wind region, allowing the losses due to drag during the sunward turn to be lowered. 
The boundary between the two regions is relatively diffuse and varies considerably, depending on solar activity, Voyager 2 crossed that boundary five times \cite{Mccomas2019}. When dynamically soaring on the termination shock, an optimal trajectory might include turning off the lift-generating mechanism when crossing the boundary to coast out to the heliopause prior to the sunward turn to pick up the larger wind gradient.

\subsubsection{Slow and fast solar wind}

The solar wind speed, in the equatorial plane of the solar system, has been measured to be between 300--500~km/s. At greater elevations ($>20^\circ-30^\circ$), the wind speed increases to 750~km/s \cite{Miralles2011, verscharen2019}. These winds are referred to as the slow and fast solar wind, respectively. Similar to dynamic soaring on a shock, an optimal trajectory while soaring on the wind shear might include a coasting phase in between turns to soar on the largest wind gradient available.

\subsubsection{Planetary magnetotail}

Planetary magnetotails offer another opportunity for dynamic soaring. Outside the Earth's magnetotail, the slow solar wind flows at speeds between 300--500~km/s. While within the magnetotail the wind is close to being static. 

\begin{figure}[h]
    \centering
    \includegraphics[scale=0.6]{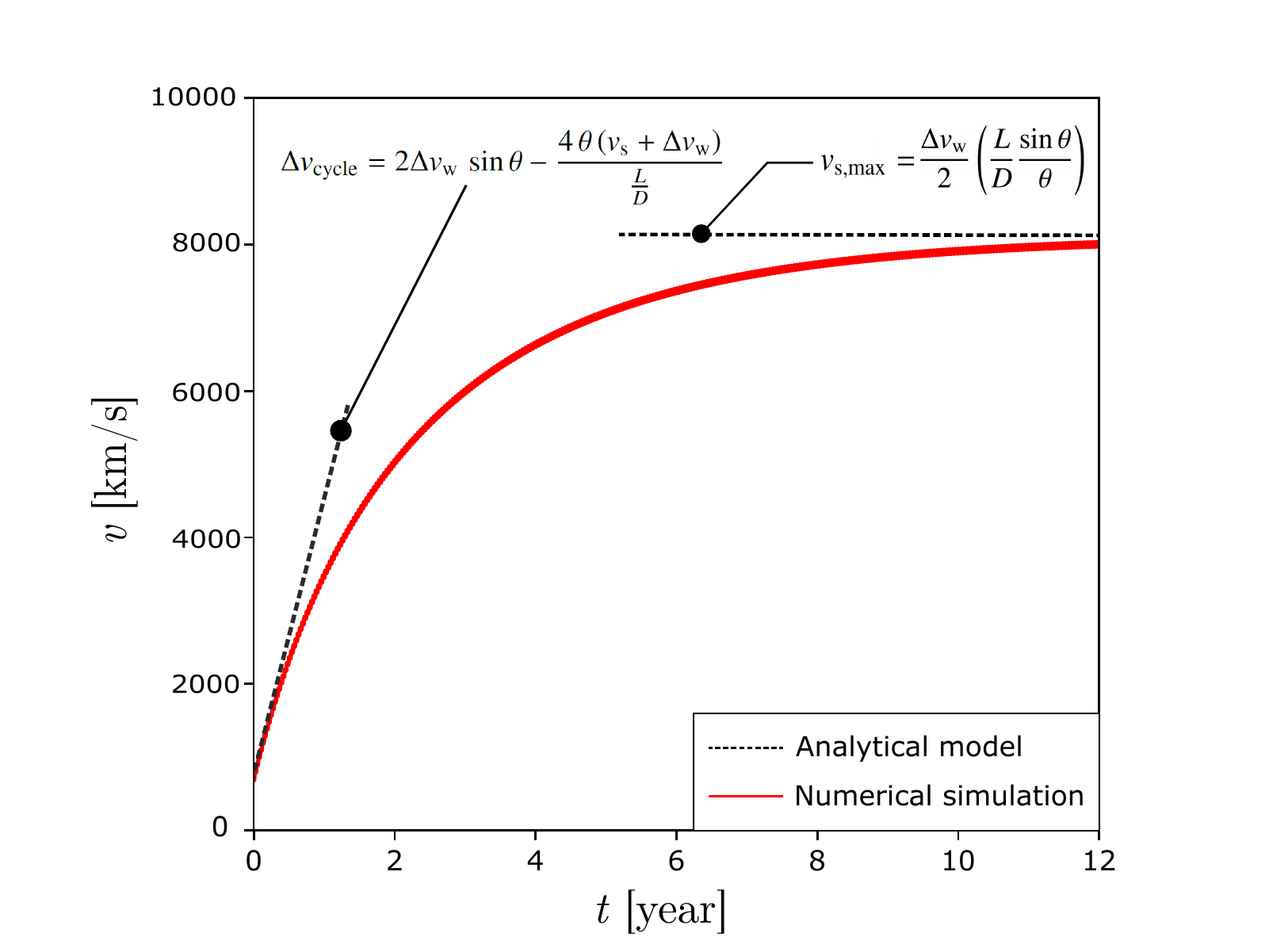}
    \caption{Spacecraft speed as function of time for dynamic soaring along a plane shock}
    \label{fig:vlim}
\end{figure}
 
\subsubsection{Numerical simulation of the trajectories}

To assess the validity of the model described above, a numerical investigation of dynamic soaring maneuvers on the termination shock has been performed both in 2D and 3D . The equations of motion have been solved, and the results plotted in comparison to the analytical results developed in Section~3.1.
The numerical simulations were performed with a lift-to-drag ratio of $25$, a constant drag acceleration of $a_\mathrm{D}=0.2~\mathrm{m/s^2}$ and a lift acceleration of $a_\mathrm{L}=5.0~\mathrm{m/s^2}$. The justification for the values used are discussed in Section~4. Dynamic soaring on the termination shock and on the fast and slow wind were simulated, respectively in 2D and 3D. 
\newline

\subsubsection{Dynamic soaring on the termination shock}
\begin{figure}[h]
    \centering
    \includegraphics[scale=0.55]{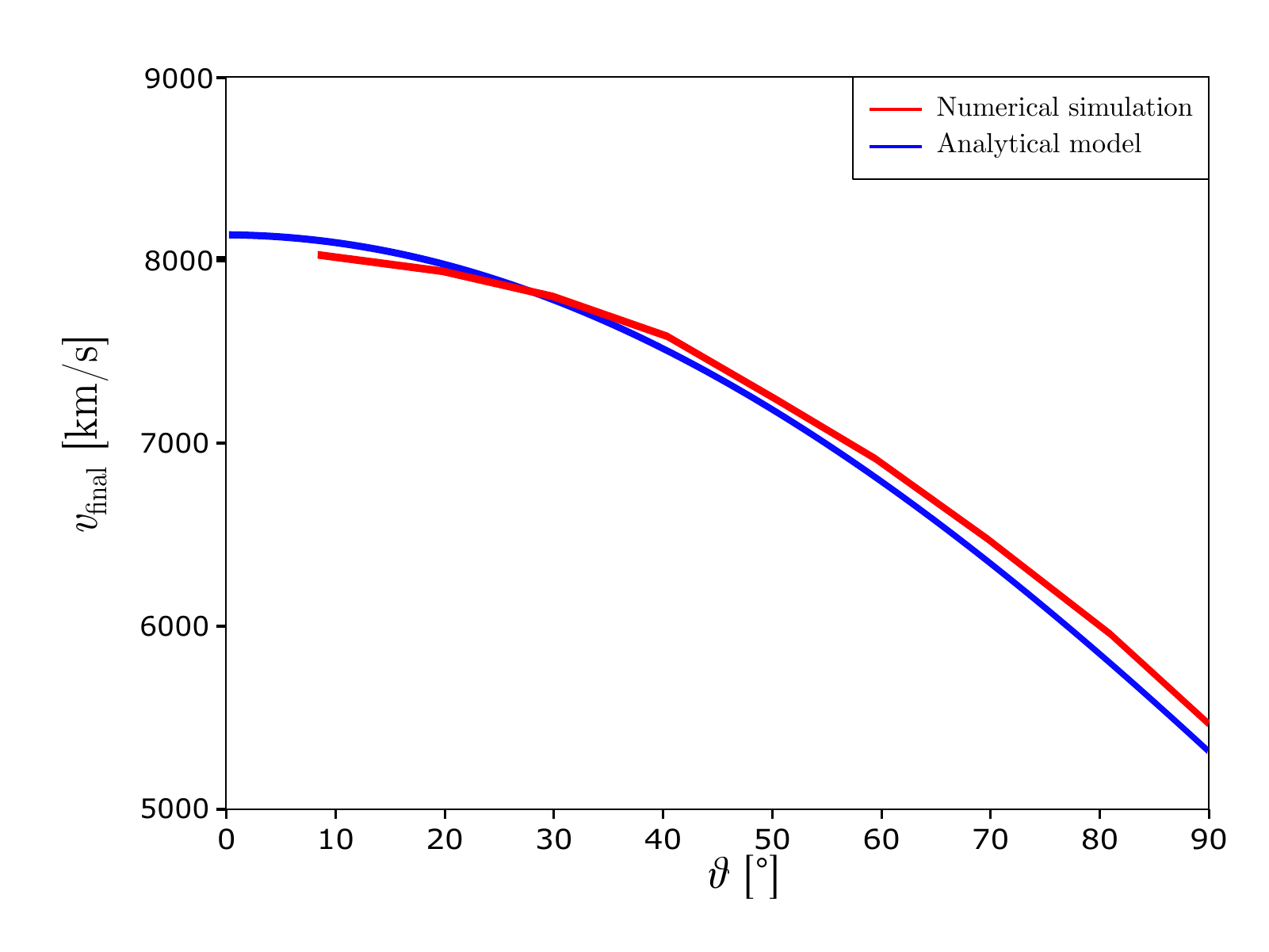}
    \caption{Maximum achievable speed on a plane shock as a function of the angle at the interface}
    \label{fig:VvsT}
\end{figure}

For the termination shock simulation, the upstream and downstream velocities are assumed to be 650 km/s and 162 km/s. In Fig.~\ref{fig:vlim} the speed of the spacecraft for dynamic soaring on a plane shock at a low value of $\theta$ is shown, corresponding to grazing along the surface of the shock. Additionally, the analytical solution of the maximum wind speed and maximum speed gain achievable bound the numerical simulation. In Fig.~\ref{fig:VvsT}, after 1.6 years of acceleration, the spacecraft reaches a velocity of $6\times10^6~\mathrm{m/s} \approx 2\%$ of $c$. 

In Fig.~\ref{fig:vlim} the maximum achievable velocity is plotted against the angle at the interface. The analytical model matches the numerical simulation within 3\%. As predicted by the analytical model, Eq.~\ref{eqn:eqatv_drag}, as the angle decreases, the losses from drag decrease. Consequently, the maximum velocity increases. 

\begin{figure}[h]
    \centering
    \includegraphics[scale=0.6]{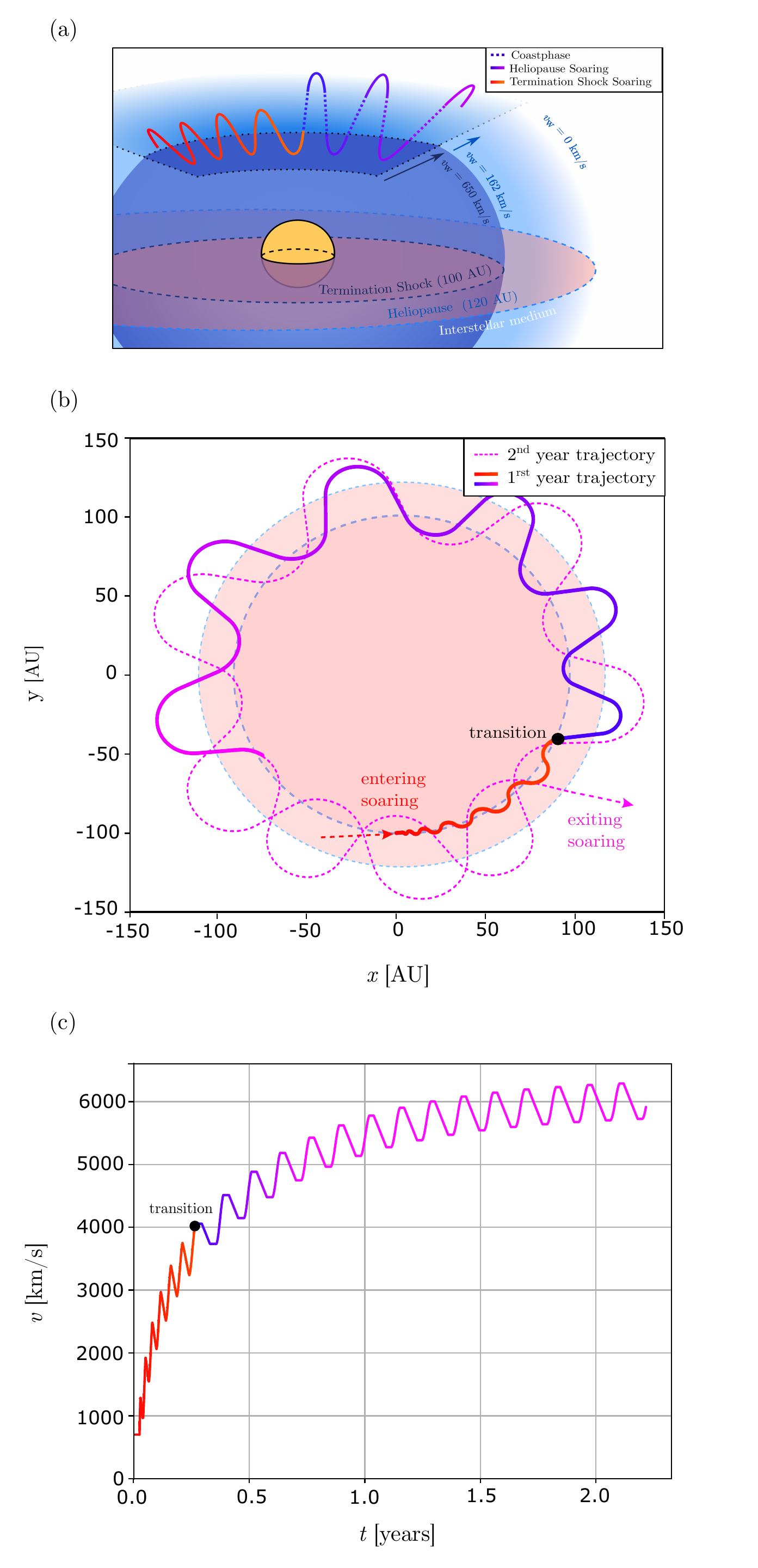}
    \caption{Curved Shock Soaring trajectory on the termination shock and the heliopause (a) schematic of the trajectory (b) numerical simulation of the trajectory projected onto the equatorial plane (c) spacecraft speed as a function of time for dynamic soaring }
    \label{fig:DSCS}
\end{figure}

In the case of a curved shock, the lift acceleration \linebreak[4] ($a_\mathrm{L}=a_\mathrm{D}~L/D$)  must be greater than the centripetal acceleration to stay in orbit on the shock.  The achievable maximum velocity reached is reduced. In Fig.~{\ref{fig:DSCS}} a numerical simulation of a spacecraft performing dynamic soaring on a forced orbit along the termination shock is shown. At 4000~km/s the spacecraft trajectory changes, instead of performing the sunward turn at the termination shock, the spacecraft coasts until the heliopause is reached to perform the turn. The wind speed at the heliopause is considerably lower than at the edge of the termination shock, allowing the spacecraft to take advantage of greater $\Delta v_\mathrm{w}$ \cite{Miralles2011}. The simulation was performed with a lift-to-drag ratio of $30$ and a constant drag acceleration of $a_\mathrm{D}=0.3~ \mathrm{m/s^2}$. The spacecraft reaches $2\%$ of $c$ after one and a half  years of acceleration.

\begin{figure}[h]
    \centering
    \includegraphics[scale=0.7]{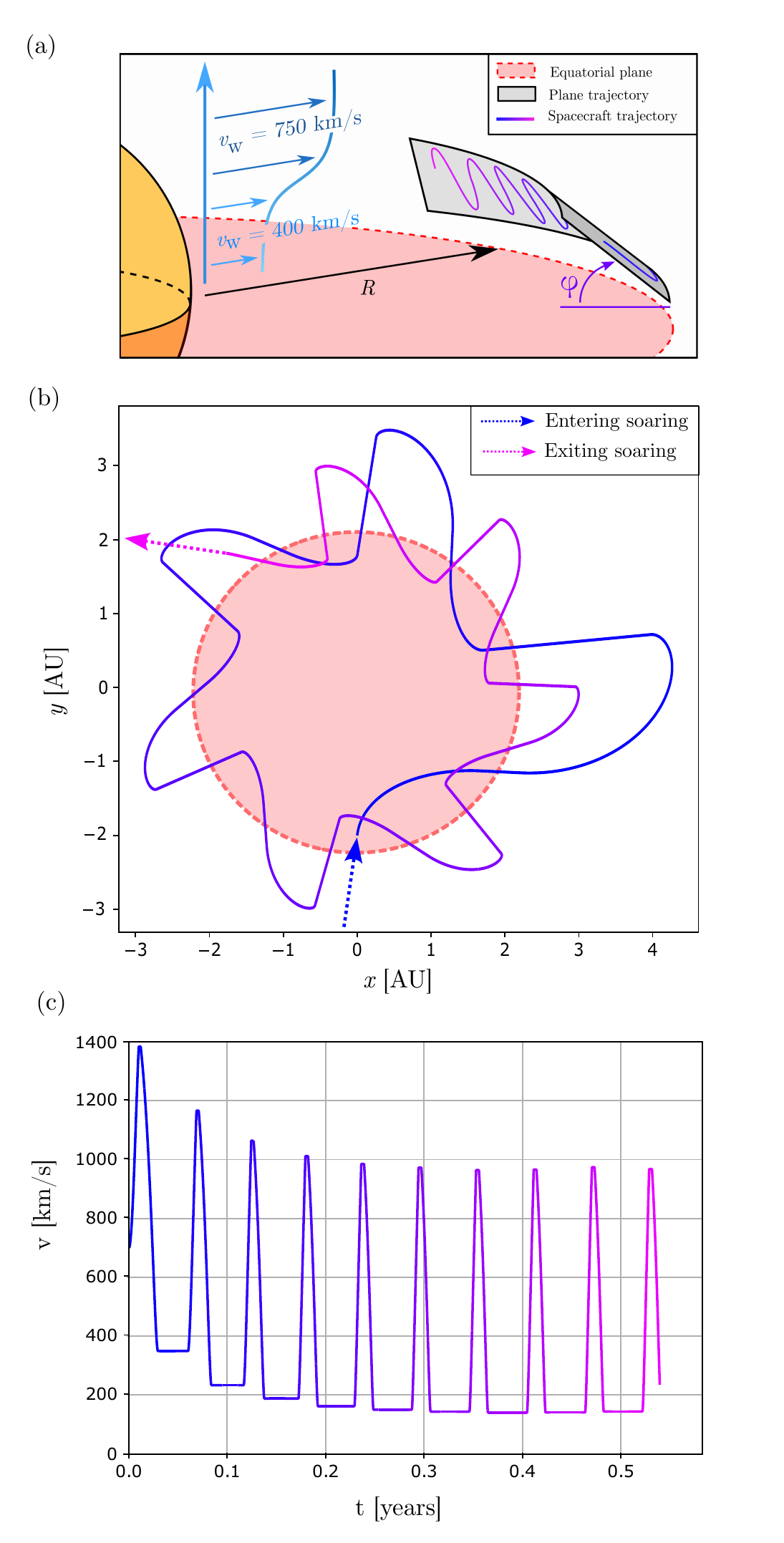}
    \caption{Three-Dimensional Dynamic Soaring trajectory on wind shear (a) schematic of the trajectory (b) numerical simulation of the trajectory projected onto the equatorial plane (c) speed of the spacecraft as a function of time}
    \label{fig:DRRW}
\end{figure}

\subsubsection{Dynamic soaring on wind shear}

For this simulation, the fast solar wind speed was taken as  750~km/s and the slow solar wind speed as 400~km/s. The trajectory of the spacecraft was constrained to a plane at an angle, $\phi$, with respect to the direction of the wind. Similar to dynamic soaring on a plane shock, as the angle $\phi$ and $\theta$ are decreased the maximum reachable velocity increases. Repeated loops above the plane of the ecliptic at a constant distance of $2~ \mathrm{AU}$ from the Sun over one month of acceleration can reach $1.4\times 10^6\approx 0.5\%$ of $c$. At low radius of gyration around the Sun, dynamic soaring is limited by the centripetal acceleration required to stay in orbit. Therefore the optimal trajectory to achieve the greatest velocity possible is to first act as a drag device to reach the velocity of the fast solar wind, and perform a single loop of dynamic soaring to reach $0.5\%$ of $c$.\\

\section{ Lift producing mechanism}

Lift is generated by accelerating the medium in the transverse direction. The low densities of the interplanetary and interstellar medium render the use of a physical wing impractical. The lift generation process must therefore accelerate large volumes of the medium. Additionally, this interaction must not decelerate the medium in the wind direction. From an aerodynamics perspective, the interaction process must not produce significant parasitic drag.

\subsection{Plasma waves}

One of the possible interaction mechanisms, fulfilling the requirements identified above, is to send plasma waves with low group velocity into the medium. The concept of launching plasma waves into the ambient interplanetary medium for propulsive purposes was previously studied by Gilland \linebreak[4] and Williams~\cite{Gilland2011}. Gilland and Williams proposed the use of Alfven waves to accelerate the medium. They showed that their design can produce net thrust in planetary magnetospheres. Sending plasma waves directly into the medium offers certain advantages. Principally, the region of the medium being accelerated is considerably larger than what any physical airfoil could achieve. Aerodynamically, this translates to a low disc loading.  
Additionally,  the medium is accelerated in the direction of propagation irrespective of its speed. The medium can therefore be accelerated perpendicularly without generating considerable parasitic drag. In other words, due to the volumetric nature of the interaction mechanism, a large region of the medium can be accelerated without significant losses. This interaction is able to produce lift independent of the orientation of the spacecraft, allowing fast modulation of the lift vector direction.  The presence of neutral particles, such as atomic hydrogen, within the plasma does not prohibit plasma waves to be launched into the medium.
However, these advantages come with the requirement that the wave must be sent unidirectionally. Producing a compact antenna design able to produce waves with, in the case of the interstellar and interplanetary medium, wavelengths on the order of kilometers is a challenging problem.  However, as will be discussed later, there already exist antenna designs capable of producing unidirectional plasma waves suitable for lift. 
As plasma waves are sensitive to the direction of the environmental magnetic field, they require the magnetic field and the propagation direction to be correctly aligned to generate lift in the desired direction.
In the different trajectories described above the spacecraft soars in gradients in wind velocity, however the termination shock and the fast-slow wind also exhibit density gradients. Consequently the frequency of the plasma wave will have to be tuned to the  local properties of the plasma to ensure the highest coupling and lift generation.

\begin{figure}[h]
    \centering
    \includegraphics[scale=0.5]{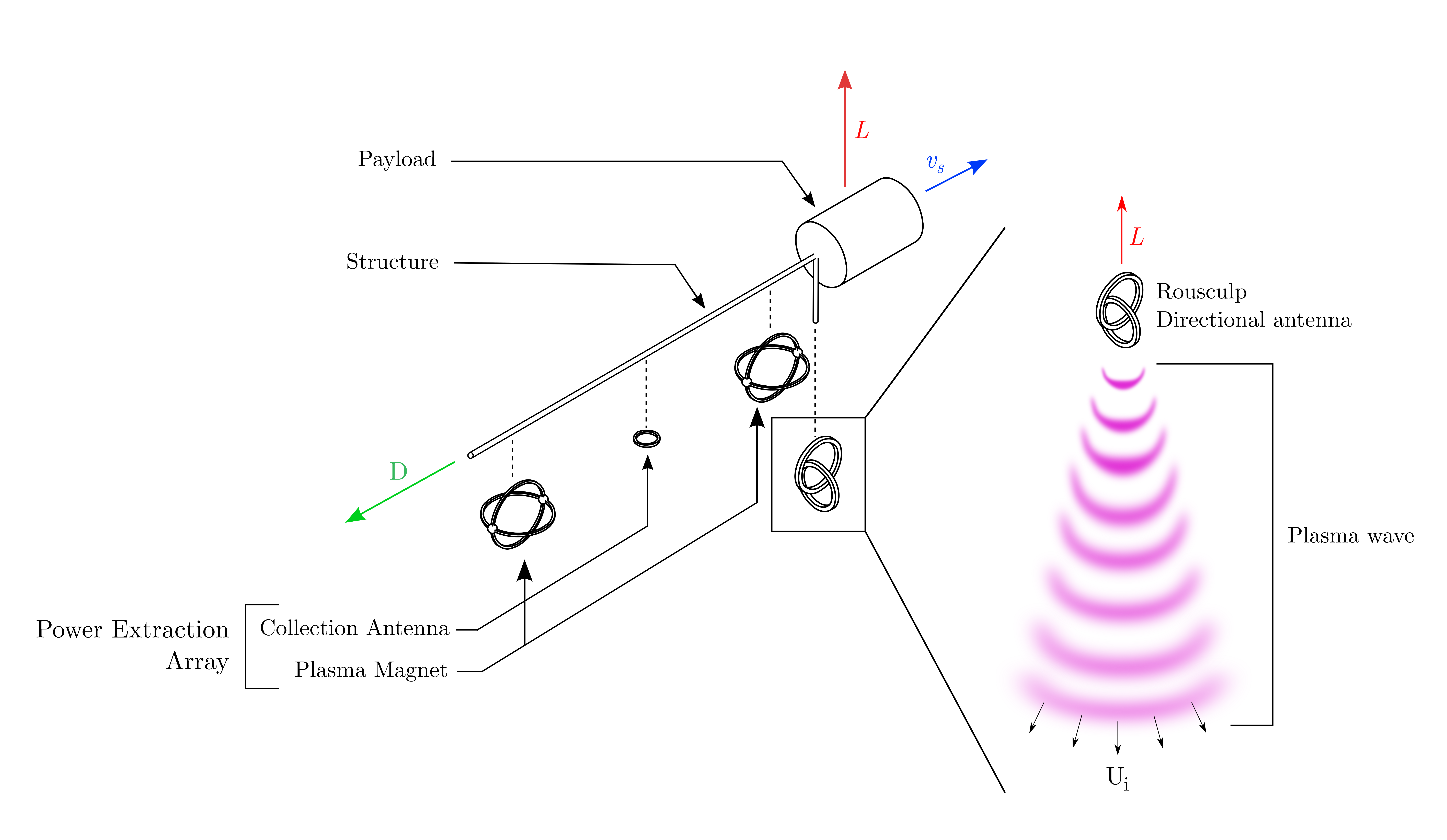}
    \caption{Schematic of spacecraft with directional plasma wave antenna that imparts momentum onto the local interplanetary or interstellar medium, generating a force on the antenna (lift)}
    \label{fig:DPWA}
\end{figure}

\subsection{Poynting vector}

The energy flux associated with an electromagnetic field is defined as the Poynting vector, $\boldsymbol{S}$.
\begin{equation} \label{eqn:eqat32}
    \boldsymbol{S}=\frac{1}{\mu}\left(\boldsymbol{E}\times \boldsymbol{H} \right)
\end{equation}
$\boldsymbol{E}$ and $\boldsymbol{H}$ refer, respectively, to the electric field vector and magnetic field vector associated with the electromagnetic wave, and $\mu$ refers to the permeability of the medium. Consider a circularly polarized electromagnetic wave travelling in the $z$ direction. 
\begin{equation} \label{eqn:eqat33}
    E(z) = \hat{x}E_{x}(z) + \hat{y}E_{y}(z) 
    = \left( \hat{x}E_{x\mathrm{0}} + \hat{y}E_{x\mathrm{0}} \right) \,\mathrm{e}^{-\mathrm{j}\beta z}
\end{equation}
\begin{equation} \label{eqn:eqat34}
    \Rightarrow 
    H(z)=\frac{1}{\eta}\hat{z}\times E(z) 
    = -\frac{1}{\eta}( \hat{y}E_{x\mathrm{0}}-\hat{x}E_{x\mathrm{0}})\,\mathrm{e}^{-\mathrm{j}\beta z}
\end{equation}
\begin{equation} \label{eqn:eqat35}
\Rightarrow
    \boldsymbol{S}=\hat{z}\frac{1}{2\eta}(|E_{x\mathrm{0}}|^2 + |E_{y\mathrm{0}}|^2)
\end{equation}
\noindent
Here, $\eta $ refers to the characteristic impedance of the medium, due to the relatively low density of the medium discussed in this analysis, vacuum properties will be assumed, $\eta_\mathrm{0} \approx377\,\Omega$. $E_{x\mathrm{0}}$ and $E_{y\mathrm{0}}$ refer to the complex amplitude of the electric field vector in the $x$ and $y$ directions, respectively. In this analysis the phase difference between the two fields is implicit.
 
\subsection{Plasma wave force}

The momentum flux associated with a wave is defined as the energy flux divided by the velocity of energy flow, i.e., the characteristic velocity of the wave. In the case of plasma waves, the velocity of energy flow is the group velocity of the wave, $v_\mathrm{g}$. Therefore, 
\begin{equation} \label{eqn:eqat35}
    \boldsymbol{P}= \frac{\boldsymbol{S}}{v_\mathrm{g}}
\end{equation}
\begin{equation} \label{eqn:eqat36}
    \Leftrightarrow \boldsymbol{F}
    = \boldsymbol{P}A_\mathrm{pw}
    = \frac{\boldsymbol{S}}{v_\mathrm{g}}A_\mathrm{pw}.
\end{equation}
From Eq. \ref{eqn:eqat36}, we can observe that the force, associated with launching plasma waves into the medium, follows the inverse of the group velocity. Equations \ref{eqn:eqat3} and \ref{eqn:eqat36} illustrate, in two different ways, why producing low group velocity is a crucial requirement for the technology, as low group velocity corresponds to low acceleration of the medium (i.e., low $\Delta v_{y}$) and, at the same time, a large lifting force. 

\subsection{Suitable plasma waves}

Numerous types of plasma waves exist, but few can produce low group velocity. Cold plasma theory offers a simple theoretical model to select suitable plasma waves. It assumes a low temperature of the electrons and ions in the plasma. The validity of this assumption will be discussed later in the paper. The group velocity is defined as $v_\mathrm{g}=\frac{\partial \omega}{\partial k}$, where $\omega$ and $k$ satisfy the dispersion relation.
Using this model, the set of suitable plasma waves can be restricted. Several  potential candidates to lift generation have been identified: the R-wave (whistler wave), the Alfven wave, the X-wave, and the magnetosonic wave.
We emphasize that these may not be the most promising implementations but rather are just offered as an example.

The R-wave and the Alfven wave both propagate in the direction of the environmental magnetic field. On the other hand, the X-wave and the magnetosonic wave propagate normal to the magnetic field. Within 10 AU of the Sun the magnetic field is quasi-dipole. At larger radius the magnetic field is primarily in the circumferential direction, perpendicular to the direction of the flow of the solar wind \cite{Owens2013, Opher2009}. Therefore a spacecraft soaring on the termination shock will have to use a wave propagating normal to the environmental magnetic field to produce the lifting force in the right direction. On the other hand, a spacecraft soaring on the fast and slow wind might need to use a combination of waves propagating in two perpendicular directions to perform the maneuver. 

\begin{figure}[h]
    \centering
    \includegraphics[scale=0.6]{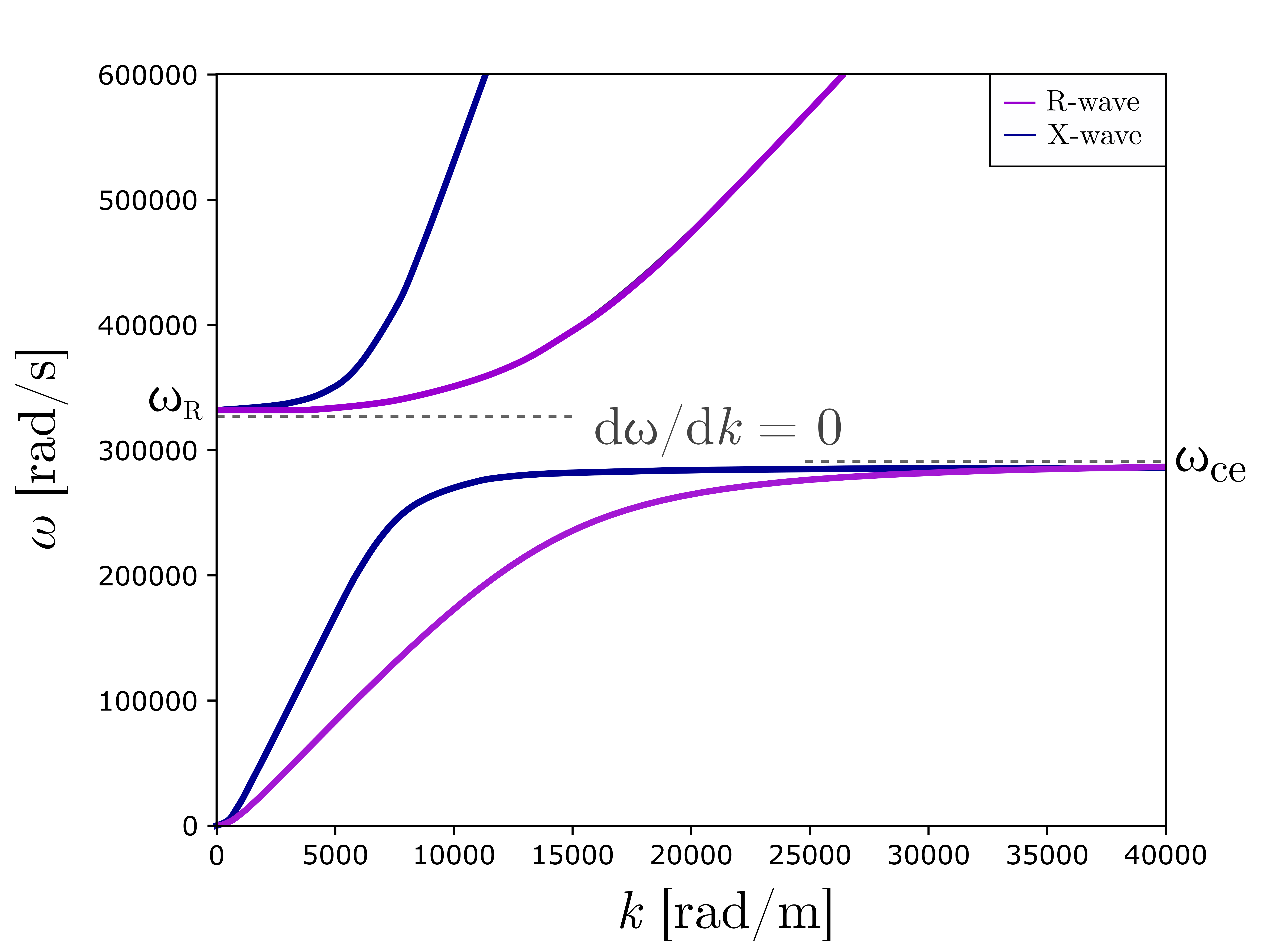}
    \caption{Dispersion relation of the R-wave and the X-wave}
    \label{fig:DRRW}
\end{figure}

As shown in Fig.~\ref{fig:DRRW}, when  $\omega$ tends to the electron gyrofrequency $\omega_\mathrm{ce}$, the R-wave and the X-wave exhibit resonating behavior. Graphically this translates to a horizontal asymptotic behavior. Therefore, as $\omega$ tends to $\omega_\mathrm{ce}$, the group velocity of the waves tends to $0$.

In cold plasma theory, the Alfven wave group velocity, $v_\mathrm{A}$, is clearly defined and is not a function of wavelength or wave vector.
\begin{equation} \label{eqn:eqat37}
    v_\mathrm{A}=\frac{B}{\sqrt{\mu_0\, \rho}}
\end{equation}
\noindent
Within cold plasma theory, MHD equations can be used to model magnetosonic waves. There exist two different modes for the magnetosonic wave: the fast and the slow mode. The fast magnetosonic wave propagates in a low $\beta$ plasma, like the solar wind, isotropically with a group velocity close to the Alfven speed. On the other hand, the slow magnetosonic wave propagates perpendicular to the magnetic field with a group velocity equal to the sound speed of the plasma. Therefore, in the interplanetary and interstellar medium, the Alfven wave and the magnetosonic wave can produce sufficiently low group velocity, $10\mathrm{-}100~\mathrm{km/s}$, to be used for lift generation.

The assumption was made earlier that the temperature of the electrons was close to absolute zero. However, thermal effects and thermal motion contribute greatly to the physics of the plasma. Therefore, the warm plasma model will be considered next. Aubry et al. \cite{Aubry1970} modeled whistler waves in warm plasmas with temperatures ranging from $120~000~\mathrm{ K}$ to $1000~\mathrm{ K}$. As thermal effects are taken into account, the behavior of the plasma becomes more complex. Due to thermal motion, the resonant behavior of the plasma observed near the gyrofrequency become pseudo-resonant behavior, i.e., the group velocity never reaches zero. From their model, the lowest group velocity achievable was determined. For the solar wind at 1~AU and an electron temperature of $120~000~\mathrm{ K}$, the group velocity for the whistler wave is $100 ~\mathrm{km/s}$. At the termination shock, with an electron temperature of $1000~\mathrm{K}$, the lowest group velocity for the whistler wave is $10~\mathrm{km/s}$. These velocities, both in the solar wind and at the termination shock, were evaluated for frequencies close to the plasma frequency, $\omega\approx1.001\,\omega_\mathrm{p}$. At $1~\mathrm{AU}$ the plasma antenna will, therefore, have to operate at 17 kHz and at 6~kHz at the termination shock. For the Alfven wave, the group velocity was derived from the observations of Shi et al. \cite{Shi2015}. The Alfven wave observed around $1~\mathrm{AU}$ had velocities of $100~\mathrm{km/s}$. This observation suggests that propagating Alfven waves at low group velocity are possible in the interplanetary medium. Due to the complexity of analyzing the coupling between the antenna and the plasma, determining exact values of maximum amplitudes, Poynting vector, and scale of the plasma antenna is left for future research.

Local properties of the plasma, such as the strength or the direction of the magnetic field, could be modified via the use of on-board magnets to modify the electron gyrofrequency and increase the coupling of the plasma antenna to the medium.
The estimated lift-to-drag ratios for these four plasma waves are given by Eq.~\ref{eqn:eqat3} with $\Delta U_{y}=v_{\mathrm{g}}$.

At $1~\mathrm{AU}$ lift-to-drag ratios of $8-15$ in the solar wind can be achieve while at the termination shock values as great as $100$ might be reached. In the dynamic soaring numerical simulations, we assumed a value of 25 for the lift-to-drag ratio at the termination shock, the discrepancy between the value used and the value given by a group velocity of $10~\mathrm{km/s}$ is due to taking into account efficiencies that arise during the different interactions between the medium and the spacecraft. 

\subsection{Plasma Antenna}

As discussed earlier, sending unidirectional waves into the plasma with a compact antenna design is not trivial and requires additional work. Launching plasma waves capable of achieving the Poynting vector magnitudes required for the trajectory described earlier requires antennae of large scales both in terms of cross-sectional area and length. Currently, we know of no compact antennae to send plasma waves unidirectionally perpendicular to the environmental magnetic field (X-wave and magnetosonic wave). However, we propose a design for the R-wave (whistler wave). We can use to our advantage one of the properties of the wave, its helicity. This property is used in Rousculp antenna design \cite{Rousculp1997}. The antennas are designed to have a preferred helicity injection direction, i.e., they couple to the plasma preferentially along one direction. In our case, this direction will correspond to the lift direction. If compact antennas capable of achieving the desired performance prove to be too complex to design, more conventional antennae arrays could be used to launch the plasma waves with the large cross-sectional areas required.

\section{Spacecraft and mission design}
\subsection{Spacecraft design}
The spacecraft requires two different systems to interact with the plasma of the solar wind, the first system extracts energy from the wind and the second accelerates the medium perpendicularly, as shown schematically in Fig.~\ref{fig:Intro01}. Power can be extracted from the medium by modulating a drag device. The power extraction currently envisioned for this technology is the plasma magnet array. As previously proposed for use in the q-drive \cite{Greason2019} it extracts power from the flow over the vehicle that is then used to accelerate on-board reaction mass. The plasma magnet array strokes the medium with two plasma magnets and power is extracted from the subsequent flow of particles. The advantages offered by the plasma magnet array for power extraction are still valid in our case. The area of interaction of the plasma magnet is not limited by the physical size of the coils but is in equilibrium with the environmental dynamic pressure. Consequently, the radius of interaction can reach tens of kilometers. Plasma magnets can, therefore, produce constant drag acceleration on the order of $a_\mathrm{D}=0.05-0.5~ \mathrm{m/s^2}$ \cite{Slough2005, Slough2007, Greason2022}. Additionally, the plasma magnet area of interaction is in equilibrium between the magnetic pressure generated by the coils and the dynamic pressure exerted by the flow of particles, thus the plasma magnet's area of interaction naturally contracts or expands as the dynamic pressure or density of the flow fluctuates.
Discussing the scale and the mass of the system was judged premature, as the scale of crucial subsystems remain undetermined and a function of the exact technology implemented for the energy extraction and the lift generation.

\subsection{Mission design}

Lift generation, by sending plasma waves into the surrounding medium, enables new types of missions. The high lift-to-drag ratio allows a variety of fast transit missions to be imagined in the inner and outer solar system. 
The system can also be used as a braking system. Transfer to an orbit about a target planet can be easily achieved by doing non-atmospheric deceleration.  In other words, by pointing the lift vector in the right direction the circumferential speed can be greatly decreased without any aerocapture or propellant burn.
To explore the possibility to achieve the greatest speed, the following architecture is proposed (see Table~1).

\begin{table}
\centering
\caption{Mission profile}
\label{tab:MissionProfile}
\begin{tabular}{@{}lrl@{}}
\toprule
Event&Time\\
\hline
Spacecraft exits Earth's magnetosphere  & T+0~month\\
Spacecraft deploys one plasma magnet    & T+0~month\\
Spacecraft reaches solar wind speed     & T+1~month\\
Arrive at termination shock             & T+13~month\\
Start of the dynamic soaring  maneuver  & T+13~month\\
Spacecraft reaches $2\%$ of $c$         & T+30~month\\

\hline
\end{tabular}
\end{table}%

Since dynamic soaring can deliver payloads to 2\% of $c$, essentially ``for free’’ (meaning, without expenditure of propellant or significant power), this technique is ideally suited to deliver reaction mass that can be used for additional stages of propulsion capable of even greater speeds. The q-drive \cite{Greason2019} is a technique by which reaction mass onboard a spacecraft can be expended via an interaction with the interstellar medium in order to concentrate the kinetic energy of the mass into the final (dry) mass of the spacecraft. The q-drive becomes particularly promising at speeds starting at 5\% of $c$, and may be able to reach speeds of 20\% of $c$. In order to span the region from 2\% to 5\% of $c$, the present authors have recently proposed the concept of \emph{wind--pellet shear sailing} \cite{Greason2022}, in which a spacecraft overtakes pellets traveling slower than the spacecraft while again interacting with the ambient medium in order to concentrate the pellet energy into the spacecraft. Both of these concepts (the q-drive and wind--pellet shear sailing) require pre-accelerating a significant reaction mass in order to work. The ability of dynamic soaring to provide the acceleration of large masses to 2\% of $c$ appears to make it ideally suited as a first stage required for these other concepts.

\subsection{Technology Roadmap}

Development of the concept of interacting with the solar wind as a means of propulsion will require experimental validation in stages, the first of which would be demonstration of significant drag against the solar wind using a magnetic structure for propulsion. The plasma magnet appears to be the highest performing in terms of accelerations of the drag concepts reviewed in the Introduction, so a plasma magnet technology demonstration would appear to be the next logical step. A recent study has proposed a small, 16U cubesat demonstrator concept termed \emph{Jupiter Observing Velocity Experiment (JOVE)} that could transit the orbit of Jupiter just six months after launch from Earth \cite{Freeze2022}. Another application of the wind-riding plasma magnet technology would be a demonstration of rapid access to the solar gravitational lens (SGL) distance ($> 550~\mathrm{AU}$). The study, called \emph{Wind Rider Pathfinder Mission}, has shown the SGL region could be accessed in less than 7 years from launch using this technology \cite{freeze2021wind}. These groundbreaking  missions would provide validation that meaningful propulsive power could be extracted from the solar wind, providing a foundation for the more advanced concept of extracting electrical power from the wind for lift-generation.

\section{Conclusions}
The ability to generate large values of lift-to-drag ratio via interaction with the flow of interplanetary and interstellar medium over a spacecraft is found to be feasible, at least from the perspective of the physical principals involved. Plasma waves near the resonant frequency for the plasma are found to be an effective means to impart a momentum change onto the flow in the transverse direction using a compact, directional antenna. Several waves are identified as having the low group velocities necessary to give effective coupling that imparts momentum change to the flow in the transverse direction; other plasma waves should also be examined. The values of lift-to-drag ratio that may be achieved ($\frac{L}{D} > 10$) are sufficiently great that a dynamic soaring maneuver can be performed by the spacecraft, enabling the vehicle to reach multiples of the difference in wind speed between different regions of the heliosphere and exceed the solar wind speed. The dynamic soaring technique appears feasible for a vehicle to achieve velocities approaching 2\% of $c$ after a year and a half by soaring along the termination shock and heliopause or 0.5\% of $c$ after one month by soaring along the slow and fast solar wind. Other structures in the heliosphere may offer even greater wind gradients, potentially providing even more significant velocity gains.

\section{Acknowledgments}
Thanks to the Interstellar Research Group and the TVIW conference which brought the authors together and led to the development of this concept. Max Greason provided some illustrations. The authors thank Christopher Limbach and Adam Crowl for useful comments and input. The authors thank Dmytro Yakymenko for feedback on the manuscript. M.N.L. was supported by the Natural Sciences and Engineering Research Council of Canada (NSERC) and the McGill Summer Undergraduate Research in Engineering program.

\section*{Nomenclature}

\begin{table}[ht]
\centering
\caption{Nomenclature}
\label{tab:MissionProfile}
\begin{tabular}{@{}lll@{}}
\toprule
Latin Symbols \\
\hline
{$A_\mathrm{pw}$}&{area of plasma wave}\\
{$a_\mathrm{D}$}&{drag acceleration}\\
{$B$}&{magnetic field strength}\\
{$c$}&{speed of light}\\
{$C_\mathrm{D}$}&{drag coefficient}\\
{$C_\mathrm{L}$}&{lift coefficient}\\
{$D$}&{drag force}\\
{$E_\mathrm{kin}$}&{kinetic energy of the spacecraft}\\
{$\boldsymbol{E}$}&{electric field vector}\\
{$\boldsymbol{H}$}&{magnetic field vector}\\
{$\mathrm{j}$}&{imaginary unit}\\
{$k$}&{angular wave number}\\
{$L$}&{lift force}\\
{$m$}&{spacecraft mass}\\
{$\dot{m}$}&{particle mass flux}\\
{$P$}&{power}\\
{$\boldsymbol{S}$}&{Poynting vector}\\
{$U, U_\mathrm{w}$}&{wind speed as viewed from the spacecraft reference frame}\\
{$v, v_\mathrm{s}$}&{spacecraft speed as viewed from the fixed reference frame}\\
{$v_\infty$}&{wind speed as viewed from the fixed reference frame}\\
{$v_i$}&{velocity change across a rotor disk}\\
{$\Delta v_\mathrm{w}$}&{difference in wind velocity}\\
{$v_\mathrm{g}$}&{group velocity}\\
\hline 
Greek Symbols\\
\hline
{$\beta$}&{ballistic coefficient}\\
{$\eta$}&{efficiency}\\
{$\eta_\mathrm{0}$}&{plasma characteristic impedance}\\
{$\mu$}&{permeability}\\
{$\theta$}&{angle at the interface}\\
{$\phi$}&{angle between the wind gradient and the spacecraft velocity}\\
{$\rho_\infty$}&{wind density}\\
{$\rho_\mathrm{p}$}&{plasma density}\\
{$\omega$}&{angular frequency}\\
{$\omega_\mathrm{p}$}&{plasma frequency}\\
{$\omega_\mathrm{ce}$}&{electron gyrofrequency}\\

\hline
\end{tabular}
\end{table}%

\newpage

\bibliographystyle{habbrv}  
\bibliography{references}

\begin{thebibliography}{10}
\expandafter\ifx\csname url\endcsname\relax
  \def\url#1{\texttt{#1}}\fi
\expandafter\ifx\csname doi\endcsname\relax
  \def\doi#1{\burlalt{doi:#1}{http://dx.doi.org/#1}}\fi
\expandafter\ifx\csname urlprefix\endcsname\relax\def\urlprefix{URL }\fi
\expandafter\ifx\csname href\endcsname\relax
  \def\href#1#2{#2}\fi
\expandafter\ifx\csname burlalt\endcsname\relax
  \def\burlalt#1#2{\href{#2}{#1}}\fi

\bibitem{Anderson1991}
J.~D. Anderson~Jr.
\newblock {\em Fundamentals of Aerodynamics}.
\newblock McGraw-Hill Education, 1991.

\bibitem{Aubry1970}
M.~Aubry, J.~Bitoun, and P.~Graff.
\newblock Propagation and group velocity in a warm magnetoplasma.
\newblock {\em Radio Science}, 5(3):635--645, 1970.
\newblock \doi{https://doi.org/10.1029/RS005i003p00635}.

\bibitem{balogh2013}
A.~Balogh and R.~A. Treumann.
\newblock The heliospheric termination shock.
\newblock {\em Physics of Collisionless Shocks}, pages 463--494, 2013.
\newblock \doi{https://doi.org/10.1007/978-1-4614-6099-2_11}.

\bibitem{Birch1989}
P.~Birch.
\newblock Dynamic compression members.
\newblock {\em Journal of the British Interplanetary Society}, 42:501--508,
  1989.
\newblock
  \urlprefix\url{https://www.orionsarm.com/fm_store/DynamicCompressionMembers.pdf}.

\bibitem{Blain2021}
L.~Blain.
\newblock World's fastest {RC} aircraft hits a stunning 548 mph---without a
  motor.
\newblock {\em New Atlas}, Jan 2021.
\newblock
  \urlprefix\url{https://newatlas.com/aircraft/dynamic-soaring-speed-record-spencer-lisenby/}.

\bibitem{Rousculp1997}
R.~{C. Rousculp}.
\newblock Helicity injection by knotted antennas into electron
  magnetohydrodynamical plasmas.
\newblock {\em Physical Review Letters}, 79:837--840, 08 1997.
\newblock \doi{10.1103/PhysRevLett.79.837}.

\bibitem{Davoyan2021}
A.~R. Davoyan, J.~N. Munday, N.~Tabiryan, G.~A. Swartzlander, and L.~Johnson.
\newblock Photonic materials for interstellar solar sailing.
\newblock {\em Optica}, 8(5):722--734, May 2021.
\newblock \doi{10.1364/OPTICA.417007}.

\bibitem{Djojodihardjo2018}
H.~Djojodihardjo.
\newblock Review of solar magnetic sailing configurations for space travel.
\newblock {\em Advances in Astronautics Science and Technology}, 1(2):207--219,
  2018.
\newblock \doi{10.1007/s42423-018-0022-4}.

\bibitem{Fermi1949}
E.~Fermi.
\newblock On the origin of the cosmic radiation.
\newblock {\em Phys. Rev.}, 75:1169--1174, Apr 1949.
\newblock \doi{10.1103/PhysRev.75.1169}.

\bibitem{Fermi1954}
E.~Fermi.
\newblock Galactic magnetic field and the origin of cosmic radiation.
\newblock {\em Astrophysical Journal}, 119:1--6, 1954.
\newblock \urlprefix\url{https://adsabs.harvard.edu/pdf/1954ApJ...119....1F}.

\bibitem{freeze2021wind}
B.~Freeze, J.~Greason, M.~Lamontagne, D.~Conway, J.~Fuller, R.~Nader, E.~Davis,
  J.~Cassibry, S.~Thomas, J.~Febres, et~al.
\newblock Wind {R}ider {P}athfinder {M}ission to {T}rappist-1 solar
  gravitational lens focal region in 8 years.
\newblock In {\em AGU Fall Meeting Abstracts}, volume 2021, pages SH15F--2081,
  2021.
\newblock
  \urlprefix\url{https://www.researchgate.net/publication/358914126_Wind_Rider_Pathfinder_Mission_to_Trappist-1_Solar_Gravitational_Lens_Focal_Region_in_8_Years}.

\bibitem{Freeze2022}
B.~Freeze, J.~Greason, R.~Nader, J.~J. Febres, A.~Chaves-Jiminez,
  M.~Lamontagne, S.~Thomas, J.~Cassibry, J.~Fuller, E.~Davis, et~al.
\newblock Jupiter {O}bserving {V}elocity {E}xperiment ({JOVE}): {I}ntroduction
  to {W}ind {R}ider solar electric propulsion demonstrator and science
  objectives.
\newblock {\em Publications of the Astronomical Society of the Pacific},
  134(1032):023001, 2022.
\newblock \doi{https://doi.org/10.1088/1538-3873/ac4812}.

\bibitem{gao2015}
X.-Z. Gao, Z.-X. Hou, Z.~Guo, and X.-Q. Chen.
\newblock Energy extraction from wind shear: reviews of dynamic soaring.
\newblock {\em Proceedings of the Institution of Mechanical Engineers, Part G:
  Journal of Aerospace Engineering}, 229(12):2336--2348, 2015.
\newblock \doi{https://doi.org/10.1177\%2F0954410015572267}.

\bibitem{Gilland2011}
J.~H. Gilland and G.~J. Williams.
\newblock The potential for ambient plasma wave propulsion: Final report to the
  \uppercase{NASA} \uppercase{I}nstitute for \uppercase{A}dvanced.
\newblock Technical report, 2011.
\newblock
  \urlprefix\url{https://www.nasa.gov/sites/default/files/atoms/files/01-niac_2011_phasei_gilland_thepotentialforambientplasma_tagged.pdf}.

\bibitem{Greason2019}
J.~Greason.
\newblock A reaction drive powered by external dynamic pressure.
\newblock {\em Journal of the British Interplanetary Society}, 72:146–152,
  2019.
\newblock
  \urlprefix\url{https://tauzero.aero/wp-content/uploads/JBIS-May-2019-Greason.pdf}.

\bibitem{Greason2022}
J.~K. Greason, D.~Yakymenko, M.~N. Larrouturou, and A.~J. Higgins.
\newblock Wind–pellet shear sailing.
\newblock {\em Acta Astronautica}, 197:408--417, 2022.
\newblock \doi{https://doi.org/10.1016/j.actaastro.2022.04.021}.

\bibitem{Heller2020}
R.~Heller, G.~Anglada-Escud{\'e}, M.~Hippke, and P.~Kervella.
\newblock {\em Astronomy \& Astrophysics}, 641:A45, 2020.
\newblock \doi{10.1051/0004-6361/202038687}.

\bibitem{Janhunen2004}
P.~Janhunen.
\newblock Electric sail for spacecraft propulsion.
\newblock {\em Journal of Propulsion and Power}, 20(4):763--764, 2004.
\newblock \doi{10.2514/1.8580}.

\bibitem{krimigis2011}
S.~M. Krimigis, E.~C. Roelof, R.~B. Decker, and M.~E. Hill.
\newblock Zero outward flow velocity for plasma in a heliosheath transition
  layer.
\newblock {\em Nature}, 474(7351):359--361, 2011.
\newblock \doi{https://doi.org/10.1038/nature10115}.

\bibitem{Lingam2020}
M.~Lingam and A.~Loeb.
\newblock Propulsion of spacecraft to relativistic speeds using natural
  astrophysical sources.
\newblock {\em The Astrophysical Journal}, 894(1):36, may 2020.
\newblock \doi{10.3847/1538-4357/ab7dc7}.

\bibitem{Lisenby2017}
S.~Lisenby.
\newblock Dynamic soaring.
\newblock 2017.
\newblock \urlprefix\url{https://youtu.be/nv7-YM4wno8}.

\bibitem{Mccomas2019}
D.~McComas, J.~Rankin, N.~Schwadron, and P.~Swaczyna.
\newblock Termination shock measured by {V}oyagers and {IBEX}.
\newblock {\em The Astrophysical Journal}, 884(2):145, 2019.
\newblock \doi{http://dx.doi.org/10.3847/1538-4357/ab441a}.

\bibitem{Giovanni2008}
G.~Mengali, A.~Quarta, and P.~Janhunen.
\newblock Electric sail performance analysis.
\newblock {\em Journal of Spacecraft and Rockets}, 45(1):122--129, 2008.
\newblock \doi{10.2514/1.31769}.

\bibitem{Millis2010}
M.~Millis.
\newblock First interstellar missions, considering energy and incessant
  obsolescence.
\newblock {\em Journal of the British Interplanetary Society}, 63:434--443,
  2010.
\newblock
  \urlprefix\url{https://www.researchgate.net/profile/Marc-Millis/publication/280742055_First_interstellar_missions_considering_energy_and_incessant_obsolescence/links/55c5164208aea2d9bdc39806/First-interstellar-missions-considering-energy-and-incessant-obsolescence.pdf}.

\bibitem{Miralles2011}
M.~P. Miralles and J.~S. Almeida.
\newblock {\em The Sun, the Solar Wind, and the Heliosphere}, pages 3--8.
\newblock Springer Netherlands, Dordrecht, 2011.
\newblock \doi{10.1007/978-90-481-9787-3_1}.

\bibitem{Opher2009}
M.~Opher, F.~Alouani-Bibi, G.~Toth, J.~Richardson, V.~Izmodenov, and
  T.~Gombosi.
\newblock A strong, highly-tilted interstellar magnetic field near the solar
  system.
\newblock {\em Nature}, 462:1036--8, 12 2009.
\newblock \doi{10.1038/nature08567}.

\bibitem{Owens2013}
M.~J. Owens and R.~J. Forsyth.
\newblock The heliospheric magnetic field.
\newblock {\em Living Reviews in Solar Physics}, 10(1):1--52, 2013.
\newblock \doi{https://doi.org/10.12942/lrsp-2013-5}.

\bibitem{Perakis2020}
N.~Perakis.
\newblock Maneuvering through solar wind using magnetic sails.
\newblock {\em Acta Astronautica}, 177:122--132, 2020.
\newblock \doi{https://doi.org/10.1016/j.actaastro.2020.07.029}.

\bibitem{Rayleigh1883}
J.~W.~S. Rayleigh.
\newblock The soaring of birds.
\newblock {\em Nature}, 27:534–535, 1883.

\bibitem{richardson2011}
P.~L. Richardson.
\newblock How do albatrosses fly around the world without flapping their wings?
\newblock {\em Progress in Oceanography}, 88(1-4):46--58, 2011.
\newblock \doi{https://doi.org/10.1016/j.pocean.2010.08.001}.

\bibitem{Richardson2019}
P.~L. Richardson.
\newblock Leonardo da {V}inci's discovery of the dynamic soaring by birds in
  wind shear.
\newblock {\em Notes and Records: the Royal Society Journal of the History of
  Science}, 73(3):285--301, 2019.
\newblock \doi{10.1098/rsnr.2018.0024}.

\bibitem{Shi2015}
M.~Shi, C.~Xiao, Q.~Li, W.~Honggang, X.~Wang, and H.~Li.
\newblock Observations of {A}lfven and slow waves in the solar wind near 1
  {AU}.
\newblock {\em The Astrophysical Journal}, 815:122, 12 2015.
\newblock \doi{10.1088/0004-637X/815/2/122}.

\bibitem{Slavin1981}
J.~A. Slavin and R.~E. Holzer.
\newblock Solar wind flow about the terrestrial planets 1. modeling bow shock
  position and shape.
\newblock {\em Journal of Geophysical Research: Space Physics},
  86(A13):11401--11418, 1981.
\newblock \doi{https://doi.org/10.1029/JA086iA13p11401}.

\bibitem{Slough2007}
J.~Slough.
\newblock Plasma sail propulsion based on the plasma magnet.
\newblock In {\em 30th International Electric Propulsion Conference, Florence,
  Italy, IEPC-2007-15}, 2007.
\newblock \urlprefix\url{http://electricrocket.org/IEPC/IEPC-2007-015.pdf}.

\bibitem{Slough2005}
J.~Slough and L.~Giersch.
\newblock The plasma magnet.
\newblock In {\em 41st AIAA/ASME/SAE/ASEE Joint Propulsion Conference \&
  Exhibit}, page 4461, 2005.
\newblock \doi{https://doi.org/10.2514/6.2005-4461}.

\bibitem{Spreiter1980}
J.~R. Spreiter and S.~S. Stahara.
\newblock A new predictive model for determining solar wind-terrestrial planet
  interactions.
\newblock {\em Journal of Geophysical Research: Space Physics},
  85(A12):6769--6777, 1980.
\newblock \doi{https://doi.org/10.1029/JA085iA12p06769}.

\bibitem{Spurio2015}
M.~Spurio.
\newblock {\em Acceleration Mechanisms and Galactic Cosmic Ray Sources}.
\newblock Springer International Publishing, Cham, 2015.
\newblock \doi{10.1007/978-3-319-08051-2_6}.

\bibitem{usmanov2006}
A.~Usmanov and M.~Goldstein.
\newblock A three-dimensional {MHD} solar wind model with pickup protons.
\newblock {\em Journal of Geophysical Research: Space Physics}, 111(A7), 2006.
\newblock \doi{https://doi.org/10.1029/2005JA011533}.

\bibitem{verscharen2019}
D.~Verscharen, K.~G. Klein, and B.~A. Maruca.
\newblock The multi-scale nature of the solar wind.
\newblock {\em Living Reviews in Solar Physics}, 16(1):1--136, 2019.
\newblock \doi{https://doi.org/10.1007/s41116-019-0021-0}.

\bibitem{Zubrin2000}
R.~Zubrin and A.~Martin.
\newblock The magnetic sail: Final report to the \uppercase{NASA}
  \uppercase{I}nstitute for \uppercase{A}dvanced \uppercase{C}oncepts.
\newblock Technical report, 01 2000.
\newblock
  \urlprefix\url{http://www.niac.usra.edu/files/studies/final_report/320Zubrin.pdf}.

\bibitem{Zubrin1991}
R.~M. Zubrin and D.~G. Andrews.
\newblock Magnetic sails and interplanetary travel.
\newblock {\em Journal of Spacecraft and Rockets}, 28(2):197--203, 1991.
\newblock \doi{10.2514/3.26230}.

\bibitem{Zwan1976}
B.~J. Zwan and R.~A. Wolf.
\newblock Depletion of solar wind plasma near a planetary boundary.
\newblock {\em Journal of Geophysical Research}, 81(10):1636--1648, 1976.
\newblock \doi{https://doi.org/10.1029/JA081i010p01636}.

\end{thebibliography}

\end{document}